\documentstyle[12pt,epsfig,rotating]{article}

\begin{document}
\begin{titlepage}
\begin{flushleft}
%
%===> Change report numbers and date
%
{\tt DESY 98-054    \hfill    ISSN 0418-9833} \\
May 1998\\
Revised December 1998\\
\vspace*{0.9cm}
\end{flushleft}

\vspace*{3.cm}
\begin{center}
\begin{Large}
{\bf  The Electronics of the H1 Lead/Scintillating-Fibre Calorimeters}  \\
\vspace*{2.cm}
H1 SpaCal Group \\
\end{Large}
%================================abstract=============================  ==
\vspace*{4.cm}
{\bf Abstract:}
\begin{quotation}
%
%===> Abstract 

The electronic system developed for the SpaCal lead/scintillating-fibre
calorimeters
of the H1 detector in operation at the HERA ep collider
is described in detail and  the performance achieved during
H1 data-taking is presented. The 10 {\rm MHz} bunch crossing rate of HERA 
puts severe constraints on the requirements of the electronics. 
The energy and time readout are performed respectively with a 14-bit
dynamic range and with a resolution of $\sim$0.4 {\rm ns}. 
The trigger branch consists of 
a nanosecond-resolution calorimetric time-of-flight for background 
rejection and an electron trigger based on analog `sliding windows'. The 
on-line  background rejection currently achieved is $\sim$10$^6$. 
The electron trigger  allows 
a low energy trigger threshold to be set at 
 $\sim$0.50 $\pm$ 0.08 (RMS)  {\rm GeV}
with an efficiency $\ge\,$99.9$\%$. The energy and time 
performance of the readout and trigger electronics is based
on a newly-developed low noise
($\sigma_{\rm noise}\sim$0.4 {\rm MeV}) wideband
($f \leq$200 {\rm MHz}) preamplifier located at the output of the
 photomultipliers which are used for the fibre light readout in
the $\sim$ 1 Tesla magnetic field of H1. 

\vspace*{1.cm}
 
\end{quotation}
\vfill

Submitted to Nuclear Instruments and Methods A

\cleardoublepage
\end{center}
\end{titlepage}
%======================Authorlist============================
\begin{Large} \begin{center} H1 SpaCal Group \end{center} \end{Large}
\begin{flushleft}
 R.-D.~Appuhn$^{6}$,              %DESY-PD     4/92         Appuhn
 C.~Arndt$^{6}$,                  %DESY                     Arndt
 E.~Barrelet$^{16}$,              %PARI-PD                  Barrelet
 R.~Barschke$^{6}$,               %DESY-ST     3/94         Barschke
 U.~Bassler$^{16}$,               %PARI-PD                  Bassler
 F.~Blouzon$^{16}$,               %PARI-TP                  Blouzon
 V.~Boudry$^{15}$,                %ECPL-PD                  Boudry
 F.~Brasse$^{6}$,                 %DESY     
 Ph.~Bruel$^{15}$,                %ECPL-PD                  Bruel
 D.~Bruncko$^{9}$
% ...

\vspace{1cm}
\noindent
{\it 
 $\:^1$ School of Physics and Space Research, University of Birmingham,
                             Birmingham, UK$^a$                 \\
 $\:^2$ Electronics Division, Rutherford Appleton Laboratory,
                    Chilton, Didcot, UK~$^a$    \\
 $\:^3$ Institute for Nuclear Physics, 
                    Cracow, Poland~$^{b}$     \\
 $\:^4$ Institut f\"ur Physik, Universit\"at Dortmund, Dortmund,
                                                  Germany~$^c$   \\
 $\:^5$ DSM/DAPNIA, CEA/Saclay, Gif-sur-Yvette, France      \\
 $\:^6$ DESY, Hamburg, Germany~$^c$                              \\
 $\:^7$ Institut f\"ur Hochenergiephysik, Universit\"at Heidelberg,
                                     Heidelberg, Germany~$^c$    \\
 $\:^8$ Institut f\"ur Reine und Angewandte Kernphysik, Universit\"at Kiel,
                    Kiel, Germany~$^c$          \\
 $\:^9$ Institute of Experimental Physics, Slovak Academy of
                Sciences, Ko\v{s}ice, Slovak Republic~$^d$       \\
 $ ^{10}$ Queen Mary and Westfield College, London, UK~$^a$        \\
 $ ^{11}$ Physics Department, University of Lund, 
                   Sweden~$^e$   \\
 $ ^{12}$ CPPM, Universit\'e d'Aix-Marseille II,IN2P3-CNRS,Marseille,France \\
 $ ^{13}$ Institute for Theoretical and Experimental Physics,
                                                 Moscow, Russia~$^f$ \\
 $ ^{14}$ Lebedev Physical Institute,
                   Moscow, Russia \\
 $ ^{15}$ LPNHE, Ecole Polytechnique, IN2P3-CNRS,
                             Palaiseau, France                  \\
 $ ^{16}$ LPNHE, Universit\'{e}s Paris VI and VII, IN2P3-CNRS,
                              Paris, France                     \\
 $ ^{17}$ Institute of  Physics, Czech Academy of
                    Sciences, Prague, Czech Republic~$^{d,g}$    \\
 $ ^{18}$ Nuclear Center, Charles University,
                    Prague, Czech Republic~$^{d,g}$              \\

 $ ^{\alpha}$ now at ILL, Grenoble, France \\
 $ ^{\beta}$ now at CRN, Universit\'e de Strasbourg, IN2P3-CNRS, Strasbourg, 
    France \\
 $ ^{\gamma}$ permanent address: LBL, University of California, Berkeley, 
    USA \\
}
\vspace{0.5cm}
\noindent
 $ ^*$ Corresponding author. Tel.: +33 1 6908 2583,fax: +33 1 6908 6428,\\
       e-mail: cozzika@hep.saclay.cea.fr.\\
 $ ^a$ Supported by the UK Particle Physics and Astronomy Research
       Council, and formerly by the UK Science and Engineering Research
       Council.\\
 $ ^b$ Partially supported by the Polish State Committee for Scientific
       Research, grant no. 115/E-343/SPUB/P03/002/97.\\ 
 $ ^c$Supported by the Bundesministerium f\"ur Forschung und
       Technologie, Germany
       under contract numbers 6DO57I, 6HH27I, 6HD27I and 6KI17P.         \\
 $ ^d$ Supported by the Deutsche Forschungsgemeinschaft.        \\
 $ ^e$ Supported by the Swedish Natural Science Council. \\
 $ ^f$ Supported by INTAS-International Association for the Promotion of
   Cooperation with Scientists from Independent States of the Former
   Soviet Union under Co-operation Agreement INTAS-93-0044. \\
 $ ^g$ Supported by GA \v{C}R, grant no. 202/93/2423, GA AV \v{C}R,
       grant no. 19095 and GA UK, grant no. 342.                \\

\end{flushleft}
%============================text============================  ==
%
%\newpage
%%\tableofcontents
%%\listoffigures
%%\listoftables
\newpage
%================== HERE STARTS THE TEXT       =====================
 
\section{Introduction }
\setcounter{figure}{0}
\setcounter{table}{0}
\label{intro}

The H1 detector has been in operation at the HERA ep collider at DESY since 
1992. 
Electrons of 27.5~{\rm GeV}
collide head-on with protons of 820 {\rm GeV} at a 10 {\rm MHz} bunch
crossing rate.
During the HERA winter shutdown 1994-1995, the H1 collaboration upgraded the
backward part\footnotemark \footnotetext{Backward refers to the electron 
direction.} 
of its detector  in order to extend deep inelastic ep scattering 
 measurements in the low {\sl Q$^{2}$} ($\sim$ 1 {\rm GeV$^{2}$})
%%($\sim$ \nolinebreak[4]{1 GeV$^{2}$})
 and low Bjorken {\sl x}  ($\sim$ 10$^{-5}$)  kinematic range.
 This upgrade program \cite{upgr93} was centred on the construction of two 
calorimeters 
 (electromagnetic \cite{spac96,spac97} and hadronic \cite{hadr96} wheels)
which are based on the lead/scintillating-fibre (SpaCal) 
 technology. In this paper we describe the electronics associated with both 
calorimeters and present the
performance achieved during normal data taking \cite{calor97}.

The major improvements related to the electronics,
compared to the previous backward detector \cite{bemc96}, concern the
physics trigger and the on-line background rejection. 
The study of low-{\sl x} deep inelastic scattering makes it necessary to 
measure the scattered 
electron down to very low energies at the level of 1-2 {\rm GeV} whereas
 incident 
electrons have 27.5 {\rm GeV}.
A first requirement for the 
backward physics trigger is the ability to run with a threshold value of
 1$\sim$2 {\rm GeV}.
A second requirement for the physics trigger is related to the spatial 
detection uniformity of this calorimeter, which 
has a mean value of $\sim${2$\%$} \cite{gore96},
an important feature  that must be preserved at the trigger level.
This has led to the construction of a trigger based on `sliding' analog
sums which enable recovery of the full deposited energy, regardless of the 
impact point of the particle.

The purpose of the on-line selection mentioned above 
 is to reject background events induced by collisions between the
  proton beam (820 {\rm GeV}) and residual molecules of gas present in the 
beam pipe
  upstream of the H1 detector.
 The rate of these collisions ($\sim${20 {\rm kHz}}) is a huge source of
 background triggers
  which must be reduced by a factor of~$\sim$10$^4$ ; the physics trigger 
rate, including photoproduction,
  is of the order of~$\sim${20 {\rm Hz}}.
The required background suppression is achieved by exploiting the path
length difference
between upstream proton background and physics energy depositions. 
The front face of the electromagnetic
calorimeter is located at $\sim1.5$ {\rm m} (2.0 {\rm m} for the 
hadronic part) from the interaction point,
thereby giving rise to a time-of-flight difference of 9 (12) {\rm ns}
 for the e.m. (hadronic) calorimeter. 
An on-line  time-of-flight (ToF) selection, at the nanosecond level, 
by a SpaCal calorimeter is possible due to its 
intrinsic low jitter ($\ll${0.35~{\rm ns}}) \cite{dago94}.
In contrast, in the previous calorimeter, out-of-time background event
 rejection was 
maintained using signals from an external set of scintillation counters.

In the following section, the general layout of the SpaCal electronics is
introduced, while sections 3-7 are dedicated to the detailed description
of each of the major components.

\section{Main features of the H1 SpaCal electronics}
\label{main}

\begin{figure}[htbp]
  \begin{center}
{\epsfig{file=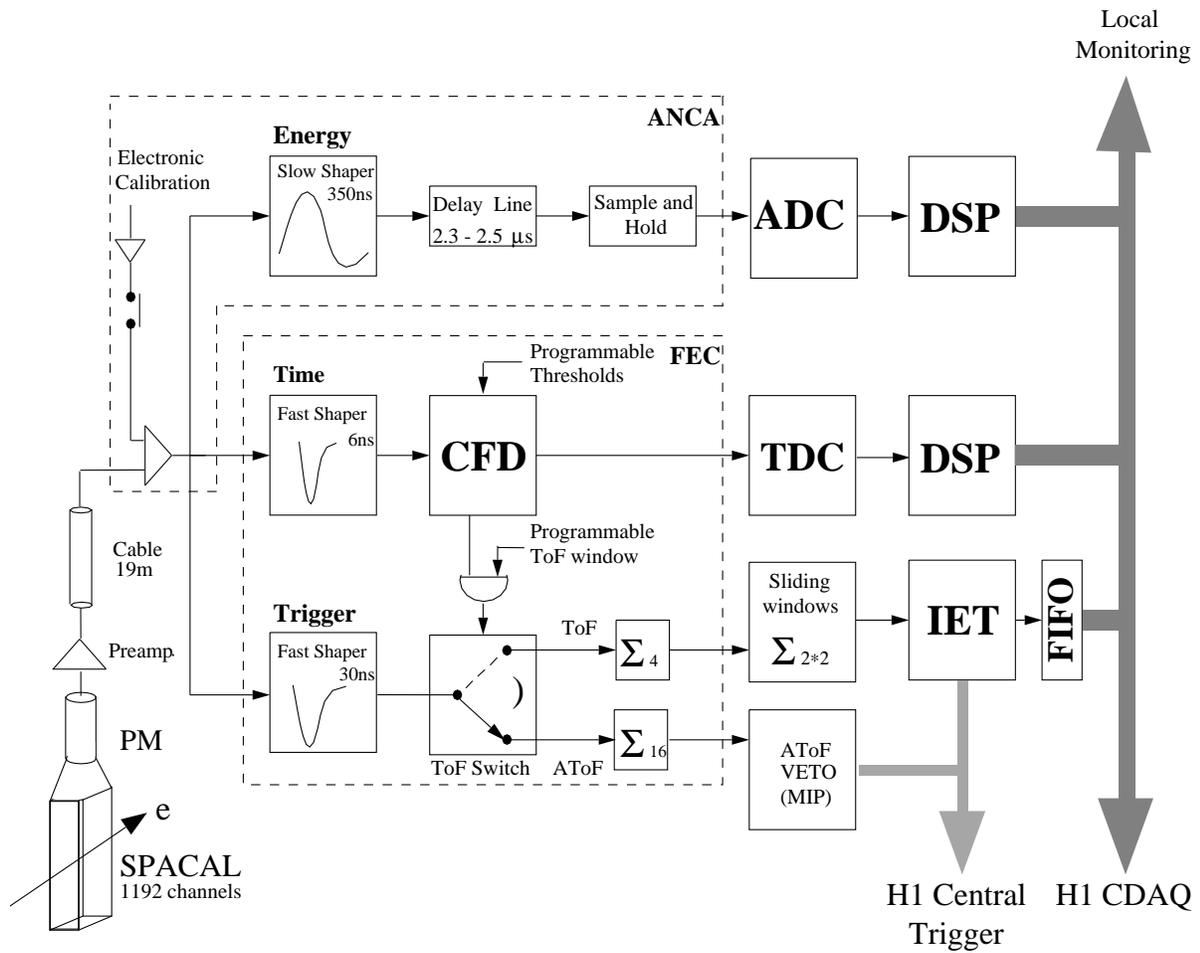,width=16.cm}}
  \end{center}
  \caption{General layout of the electronics associated to the SpaCal
 electromagnetic section.
   The signal peaking-time values are given in the shaper boxes.}
    \label{Bloc-diag}
\end{figure}

 Fig.\ref{Bloc-diag} shows the electronics associated with the 1192 
channels of the electromagnetic calorimeter. Except for the electron trigger,
the same electronics is also used for the 136 channels of the hadronic 
detector, so only minor differences will be mentioned when necessary, 
the electronics description always referring to the electromagnetic wheel.

The basis of the improvement discussed in the introduction is a very
low-noise wideband readout of the SpaCal light, achieved by a combination of a 
photomultiplier and a charge preamplifier.
The fiber light is read out by
fine-mesh Hamamatsu R5505 phototubes (R2490-05 for the hadronic section)
which provide a gain of $\sim${10}$^{4}$ ($\sim${2} $\times$ 10$^{4}$
for the hadronic section) in the $\sim$1 {\rm Tesla} magnetic field of H1. 
The detailed
characteristics and performance of these photomultipliers are given in
references \cite{spac97,phot94,spacpm}. In order to transmit the
 photomultiplier pulses to
the next stages of the electronics with the best signal to noise ratio,
a charge preamplification is performed at the anode output of the 
phototube. The wideband preamplification associated 
with the photomultiplier tube is described in the next section, where 
it is shown that a dynamic range greater than 17 bits at a frequency
of 50 {\rm MHz} is achievable.

 As shown in Fig. \ref{Bloc-diag}, at the receiver end of the 19 {\rm m} long 
50 $\Omega$ transmission cables the pulses are fed into three parallel
signal processing chains, which are:

\begin{itemize}
  \item  The {\it energy readout} branch described in Sect. \ref{energy},
  \item  The {\it timing branch} (Sect. \ref{tof});
as shown in Fig.\ref{Bloc-diag}, the constant-fraction discriminator 
(CFD) output of the timing branch is
on one hand digitized by a TDC system (Sect. \ref{tdc}) and, on the other hand,
used in coincidence with a programmable gate for the on-line time 
selection of the physics events. This function as well
as the two fast shapers and the analog sums ($\sum_{4}$ and $\sum_{16}$)
are performed on the calorimetric ToF board described in Sect. \ref{tof}.   
  \item  The {\it trigger branch} (section \ref{trigger})
which includes the level-1 physics trigger
mentioned in the introduction, and the total energy E$_{tot}$ first level
triggers for physics
events as well as for background collisions; the latter are used as vetoes
for other H1 triggers.
\end{itemize}

The three branches are characterized by different time constants, 
as can be seen 
in Fig. \ref{Bloc-diag} where peaking-time values are indicated. For on-line 
background rejection (timing branch), the short time-constant enables the 
high frequency information of the pulses to be recovered. For the trigger
branch, the shaping is chosen in order to get the minimum time occupancy 
corresponding to the period (96~{\rm ns}) of the  HERA machine clock. 
For the energy readout branch,
the time occupancy per channel due to the $\sim$20~{\rm KHz} background rate
allows a longer shaping time compared to the trigger branch.

Part of the electronics (the ANCA and FEC boards of Fig.\ref{Bloc-diag})
is located close to the iron yoke of the H1 
detector in analog boxes designed for optimal shielding and access.
Each analog box contains 8 pairs of energy read-out/calorimetric-ToF boards.
 The electronics of the trigger elements, and the TDC and ADC systems are 
located 20 {\rm m} further away, in the main H1 electronics trailer.

\section{Preamplification}
  \label{bases}

 Usually, a photomultiplier is read out by connecting 
a matched resistor R$_m$ at the end
of a transmission cable. With this configuration, the 
transimpedance gain:
\begin{equation}
 Z_t = \frac{V_{out}}{I_{pm}} = R_m
\end{equation} 
is constant up to frequencies
$f_c=1/(2\pi\tau_c)$ where $\tau_c=R_m\,.\,(C_{pm}+C_{st})$; $C_{pm}$ is the 
capacitance of the last dynode-anode of the phototube and $C_{st}$ 
is the stray capacitance. 
$V_{out}$ is the voltage measured across the resistor and $I_{pm}$ the current
pulse delivered by the photomultiplier. By using such a readout,
it is implicitly assumed that the photomultiplier gain (10$^6-10^7$)  ensures 
a high signal-to-noise ratio, independent of the signal processing performed 
in the
next stages of the electronics ; 
here noise means the pickup noise which is added to the signal 
through the transmission line, the current noise contribution of the dark 
current being negligible.

\begin{figure}[htb]
 {\epsfig{file=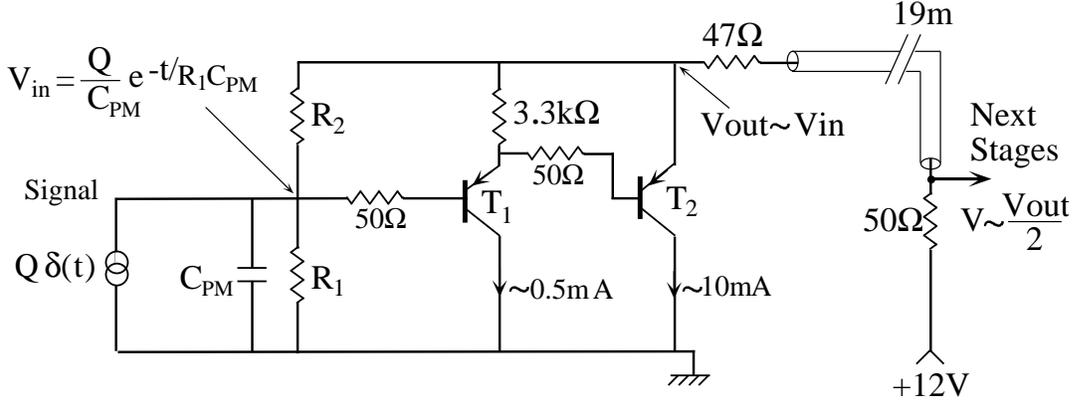,%
  width=0.9\textwidth,clip=,%
  bbllx=34pt,bblly=62pt,bburx=481pt,bbury=237pt}}
\caption{Layout of the charge preamplification.}
\label{base}                                                        
\end{figure}

However, in our present application the  lower gain 
($\sim${10$^4$}) \cite{spacpm} of the fine-mesh
phototubes inside the H1 magnetic 
field does not ensure such a  high signal-to-noise ratio.
Therefore it is necessary to
preamplify the charge  of the current pulse%
\footnote{SpaCal pulses have a rise time of $\sim${5~{\rm ns}}
and decay with a time constant of $\sim${3~{\rm ns}}.}
 $I_{pm}$~=~$Q\,\delta(t)$
delivered by the photomultiplier  (see Fig. \ref{base}),
using the dynode-anode capacitance $C_{pm}$ as the integrating
capacitor. The step voltage $Q\,{/}\,{C_{pm}}$ decays according to the time
constant $\tau_{pm}$=$R_{1}\times C_{pm}$, where the mean value of $\tau_{pm}$
has been set to 
180 {\rm  ns} ($R_{1}$~=~12~k$\Omega$, $C_{pm}\sim15~{\rm pF}$ )
in order to avoid saturation
of the electronics by pile-up effects; 
this time constant $\tau_{pm}$ varies from channel to channel
(20$\%$ maximum) due to the variation of the geometrical capacitance value 
of the phototube. For the hadronic calorimeter, the fine-mesh phototubes 
R2490-05 have
a larger capacitance value ($\sim$\,50\,{\rm pF}), so $R_1$ has been 
chosen equal to 3.3~k$\Omega$.

  The two emitter-follower transistors (see Fig. \ref{base}) drive the voltage
into the transmission cable which is matched at both ends; matching  at 
the emitter side  reduces to a negligible level reflected pulses 
 due to residual mismatching at the receiver end. For a background event,
such delayed pulses (about two bunch-crossing 
later) could lead to wrong low-energy physics triggers.

The transimpedance gain of the arrangement of Fig. \ref{base} is given by:

\begin{equation}
Z_t(s)\,=\,\frac{V_{out}}{I_{pm}}\,=\,\frac{R_1}{1+sR_1C_{pm}}\,\sim\,\frac{1}
{sC_{pm}}
\end{equation}
\noindent
with $s=\,\,j\omega$ the complex frequency. In this expression, 
the higher poles corresponding to the transition frequency 
($f_t$$\sim${5~{\rm GHz}}) of the 
transistors (Philips BFQ23) are neglected and a unit voltage gain for 
both transistors is assumed.
Comparing this transimpedance gain with the one provided by the usual readout
discussed above, (Eq. 1), we find that, for $R_m$\,=\,50~$\Omega$,
$Z_t(s)\geq50~\Omega$
over the wide frequency range $f\,\leq100$~{\rm MHz}.

There are two reasons to use such a preamplifier instead of a conventional
charge preamplifier with a feedback capacitor  $C_f$:
\begin{itemize}
 \item
  For an energy
deposit of $\sim${30}~{\rm GeV} inside a SpaCal tower, 
we  already achieve  a step voltage amplitude 
\footnote{Voltage range delivered by the preamplifier is 4~{\rm V}.}
(a light yield of 1 photoelectron per {\rm MeV} deposited energy is assumed)
$Q/C_{pm}$ of $\sim${3.2~{\rm V}}.
So there is no need to increase the gain by using a smaller integrating 
capacitor,
 \item 
   With a feedback charge preamplifier, the rise-time of the output pulse 
is related to the transimpedance $g_m$ 
of the input transistor and to the total input capacitor.
As these two quantities
are channel dependent, the use of a feedback charge preamplifier
would have introduced  timing performance variations,  
an undesirable feature for the calorimetric ToF function.
\end{itemize}

A drawback of the arrangement of Fig. \ref{base} 
is the gain dependence on the value of $C_{pm}$,
an undesirable feature which is usually removed in a conventional charge 
preamplifier 
by knowledge of the value of $C_f$. However, with phototubes this
gain dependence is compensated  when the amplification gain of the
phototubes is tuned by balancing their response in terms of energy.

The noise contribution introduced by this preamplification
is discussed below in order to  show
that, in terms of energy, it is below 1 {\rm MeV}.
The noise spectral density $S_n(s)$ at the output of the second transistor 
(Fig. \ref{base}) is
given by the following expression:

\begin{equation}
S_n(s)\,=\,e_n^2\,\oplus\,i_n^2\,Z_t^2(s)
\end{equation}
\noindent
where $\oplus$ stands for addition in quadrature ; 
$e_n$ and $i_n$ are series and
parallel noise generators respectively, 
$Z_t$ the impedance given by equation 2. 

The series noise generator is equal to:

\begin{equation}
e_n^2\,=\,4kT\,(2r_{bb'}\,\oplus\,\frac{1}{2g_{m1}}\,\oplus\,\frac{1}{2g_{m2}})
\end{equation}
\noindent
where $r_{bb'}$$\sim${10}\,$\Omega$ is the base spreading resistor of the 
transistor and $g_{m1}$\,($g_{m2}$) the transconductance of the transistor 
$T_1$\,($T_2$).
As the collector current
 $I_{c_1}$ of T${_1}$ is equal to \,$\sim$~0.5~{\rm mA},
the term $1/2g_{m1}$ amounts to $\sim${25}\,$\Omega$ 
while the contribution  $1/2g_{m2}$$\sim${1}$\Omega$
($I_{c2}$\,$\sim$~10~{\rm mA}) from T${_2}$ is negligible.
With these values of $r_{bb'}$ and $g_{m1}$, the series noise spectral 
density $e_n$\,$\sim${\,0.75}\,nV/$\sqrt{{\rm Hz}}$.

The parallel noise generator $i_n^2$ is equal to:

\begin{equation}
\label{pnoise}
i_n^2\,=\,4kT\,(\frac{1}{R_1}\,\oplus\,\frac{1}{R_2}\,)\,\oplus\,2\,q\,I_{b1}
\end{equation}

\noindent
 where $R_1$ and $R_2$ (1.8~k$\Omega$) are the resistors of the voltage
 divider (Fig. \ref{base})
 and $2qI_{b1}$ the shot noise of the base current of the transistor T$_1$;
$I_{b1}$$\sim$15~$\mu$A, the current gain
$\beta$ being equal to $\sim${30} for the BFQ23 transistor. With these values,
it is found that $i_n$\,$\sim${3.86}~pA/$\sqrt{{\rm Hz}}$, where the dominant
contribution is provided by the resistor $R_2$; in equation (\ref{pnoise}),
we have neglected the noise contribution $2qI_{b2}$ of the transistor T$_2$.

From equation (3), using the above values for $e_n$ and $i_n$, 
it can be seen that series and parallel noise contribute equally at a
frequency $f_n$$\sim$~5~{\rm MHz}, the parallel noise dominating at lower
frequencies. The noise contributions can be 
expressed in terms of  equivalent noise charge ($ENC$) 
for the three shapers of the
next stages (see Fig. \ref{Bloc-diag}) by the usual equation (ref.\cite{rad}):
\begin{equation}
\label{fnoise}
ENC\,=\,C_{pm}\,\frac{e_n\,J_a}{h_{max}\,\sqrt{\tau}}\,\oplus\,\frac{i_n\,J_b}
{h_{max}}\,\sqrt{\tau}
\end{equation}
$h_{max}$ is the maximum  of the 
preamplifier and filter response to a current pulse
$Q\delta(t)$, $J_a$ and $J_b$
are the series and parallel noise integrals.
The values of $h_{max}$,\,$J_a$ and $J_b$
vary according to the shaping performed in the next stages, which 
can be approximated by:
\begin{itemize}
 \item 
(CR)(RC) with CR$\sim${3}~{\rm ns} for the timing trigger, 
 \item
(CR)(RC)$^3$ with RC\,=\,10~{\rm ns} for the electron trigger branch, and
 \item
  (CR)(RC) with RC\,=\,180~{\rm ns} for the energy readout.
\end{itemize} 
It should be noted that equation(\ref{fnoise}) is valid for the trigger 
shapers because the approximate expression $Z_{t(s)}$~$\sim$~$1/sC_{pm}$
(equation 2)
can be used (RC$\ll\tau_{pm}$). For the readout branch, 
this approximation does not hold but the $ENC$ formula is still valid provided 
the transfer function of a (CR)$^2$(RC) shaper is used for the computation 
of $h_{max}$ and $J_b$,
the second pole being the one of the charge preamplifier;
the values of $h_{max}$,$J_a$ and $J_b$
can be found in reference (\cite{taille}).

%\input{../debut}
%\input{../com}
%\begin{document}
\begin{table}[htb]
\begin{center}
\begin{tabular}{| l || c | c | c | c | c | c | c |}
\hline
\ Electronics &\ Shaper &\ $h_{max}$ &\ $J_a^2$ &\ $J_b^2$ &\ Series &\ Parallel &\ Total \\
\ Branch &\ ($\tau$=RC) &\      &\      &\         &\  ($e^{-}$~(RMS)) &\ ($e^{-}$~(RMS)) &\ ($N_{pe}$)\\
\hline \hline 
\ Timing &\ (CR)(RC) &\ 0.367 &\ 1/8 &\ 1/8 &\ 1120 &\ 1270 &\ 0.17 \\
\ Trigger &\ $\tau \sim$ 3~ns & & & & & & \\
\hline 
\ Electron & (CR)(RC)$^3$ & 0.224 & 1/64 & 5/64 & 390 & 2990 & 0.3 \\
\ Trigger & $\tau=10$~ns & & & & & & \\
\hline
\ Energy & (CR)$^2$(RC) & 0.23 &  1/8 & 1/32 & 255 & 7790 & 0.78 \\
\ Readout & $\tau=180$~ns & & & & & & \\
\hline
\end{tabular}
\end{center}
\label{tab:noise}
\caption{ Series and parallel noise contributions (expressed in number of electrons (RMS)) of the preamplifier
for the three electronic branches. The total noise is expressed in number of 
photoelectrons ($N_{pe}$), assuming a photomultiplier gain equal to $10^4$.}
\end{table}
%\end{document}

The noise contributions of the preamplifier for the three shapers
 are summarized in table (1). From these values, one can conclude that:
\begin{itemize}
  \item
   The noise is lower than 1~{\rm MeV} for both types of shapers
  \item
   The achievable dynamic ranges
for the energy readout and timing trigger are greater than 15 and 17 bits 
respectively. It is shown in the next sections that current performance
achieved in the H1 experiment are 14 and 15 bits respectively.
\end{itemize}

%
%============================ ENERGY READ-OUT ==============================
  \section{The energy read-out chain}
  \label{energy}

\begin{figure}[htb] \centering \unitlength 1mm
  \begin{picture}(140,110)
\put(49,105){\epsfig{figure=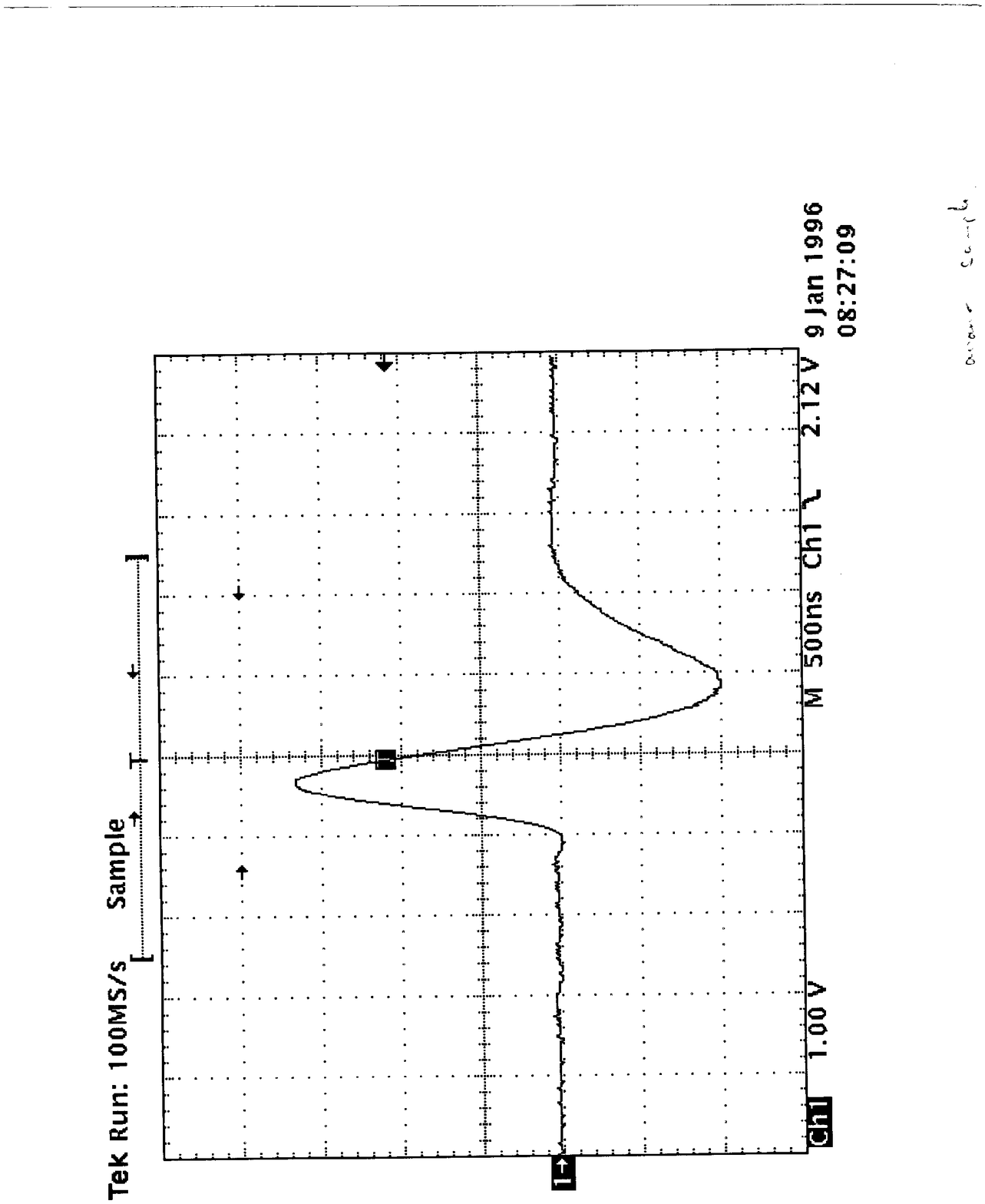,height=59 mm,angle=-91,clip=}}
 \put(-10,-10){\epsfig{figure=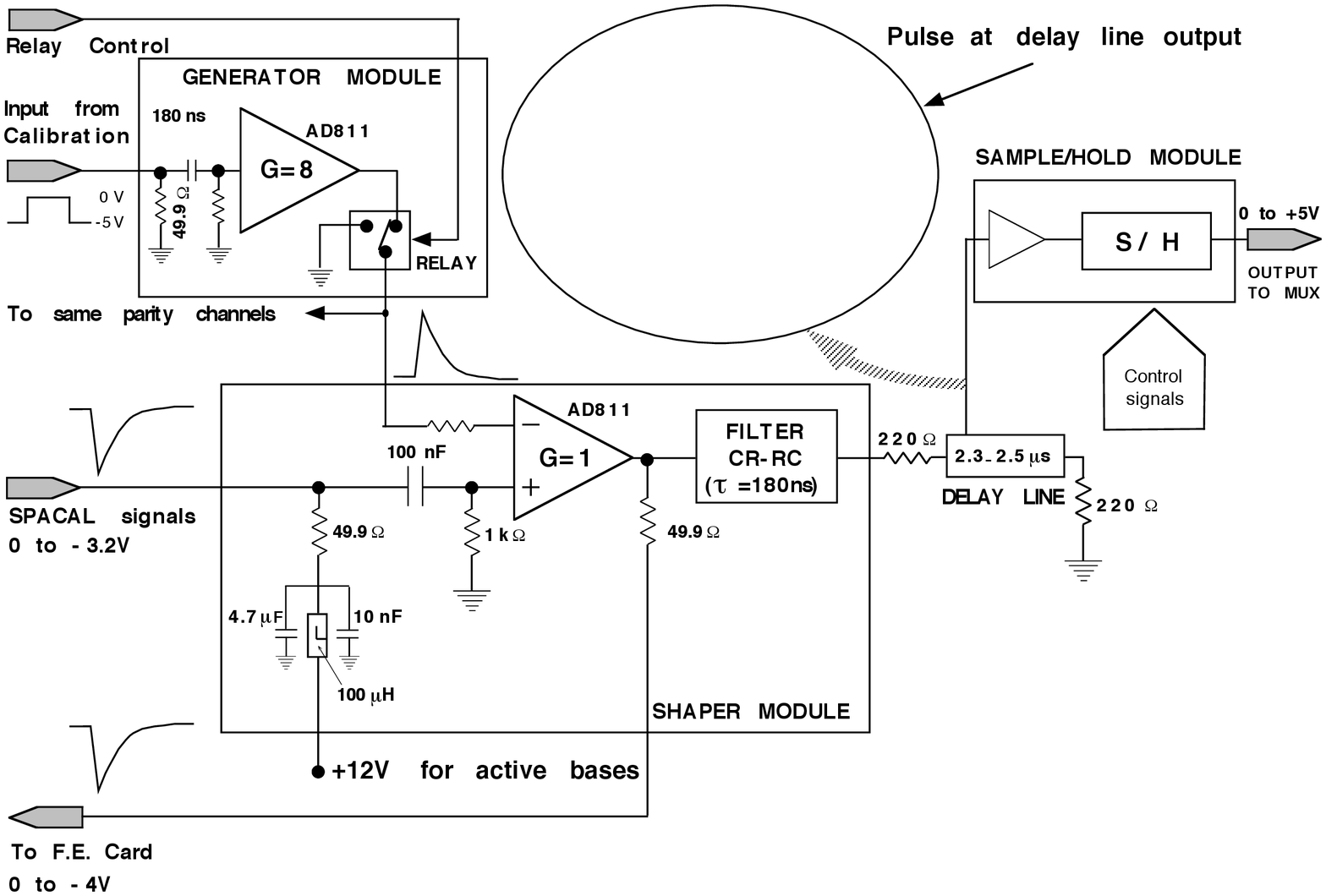,%
 height=12cm,angle=0}}
  \end{picture}
  \caption{Schematic layout of one analog card (ANCA) channel. 
The pulse at the output of
the delay line is shown with a  time scale of 0.5 $\mu$s per square.}
  \label{anca}
\end{figure}

The energy read-out is performed by 87 electronic boards  which process
 16 SpaCal channels each. As indicated in
Fig. \ref{anca} these boards (labelled ANCA in Fig. \ref{Bloc-diag}) 
supply the voltage of the preamplifiers and
the impedance termination of the 50 $\Omega$ cables.
The output of the unit-gain
Analog Devices AD 811 amplifier is sent in parallel to the calorimetric
ToF board and to the signal processing chain which consists of:

\begin{itemize}

 \item
 A (CR)(RC) filter with a time constant $\tau$~=~180~{\rm ns}. 
As the mean value of the pole of
the preamplifier R$_1$C$_{pm}$ is $\sim${180}~{\rm ns} this filtering 
is equivalent
to a (CR)$^2$~RC shaper, providing bipolar output pulses as shown in Fig. 
\ref{anca}. This bipolar shaping was chosen so as to minimize energy pile-up 
effects for the detector channels close to the beam-pipe.
 
 \item
  A delay line which has been  custom built \cite {secre} and housed in a 
shield against the
stray field ($\sim$~100~{\rm Gauss}) present close to the iron yoke.
Special care was taken in its design in order to reduce to a negligible level 
parasitic oscillations which can occur before and after the  pulse.
 A fine adjustment of the delay 
between 2.3 and 2.5 $\mu$s in steps of 20~{\rm ns}, set once by jumpers,
 enables alignment of the peak position of the bipolar pulse with 
respect to the L1keep signal sent by the H1 first level (L1)  trigger.
This peak position is measured to an accuracy of $\sim${1}~{\rm ns} by the 
electronics calibration described below.

 \item
 The sample and hold S/H (Fig.\,\ref{anca}) which stores the value of the 
 signal when an L1keep signal is received.
 The control switches are based on the same principle as
 those used in the H1 liquid argon electronics \cite {LAL_seq}
 in order to use the same sequencer,
 but the bandwidth has been increased significantly to deal with the faster 
 SpaCal signals.  
  
\end{itemize}   

The readout of the stored  amplitudes by the S/H module
  uses the same multiplexing, 
dual output gain (1 and 4), ADC and DSP  electronics
as for the H1 liquid argon calorimeter (see ref. \cite{Lar,H1nim});
the pedestal subtraction, zero suppression and gain selection
are performed by the DSP.  The dual-gain 
system was chosen so that the effective dynamic range is 14 bits with a 12 bit
ADC. A single coefficient giving the ratio of these two gains is necessary 
for the 128 channels of one analog box and is known to better than 10$^{-3}$.

 Checks, linearity measurement and delay calibration of the
bipolar pulses are performed by electronic pulses
  fed into the inverting input of the gain-1 AD811 amplifier
 (Fig.\, \ref{anca}). Square voltage  pulses received 
by the generator modules (Fig.\,\ref{anca} ) are provided by the
calibration system  developed
for the H1 liquid argon calorimeter \cite{LAL_cal}. In a generator module,
which drives 8 readout channels, the square pulse is derived by
a (CR) of 180~{\rm ns} in order to simulate SpaCal pulses. 
The gain-8 (Fig.\, \ref{anca}) is 
necessary to cover the full 4-Volt dynamic range but 
 introduces a total coherent noise of $\sim${12} MeV
 on the 8 readout channels. 
 During normal data-taking, the output
of   the generator module is disconnected by a relay in order to restore
the  minimum  noise configuration and to minimize high frequency cross-talk
~($\leq$3.5\% at 100 MHz).

The integral linearity of the full electronics chain has been found
to be of the order of 1$\%$. 
As can be seen in Fig. \ref{anca}, electronic pulses are 
also useful for checking the trigger branches.
However, for the time calibration
of calorimetric ToF boards (sect.\ref{tof}),
the square-pulse generator of the liquid argon calorimeter is replaced by a 
commercial product, HP 8082A,  which simulates better
the fast rising-edge of the SpaCal pulses. 
The  drawback is that the time calibration
procedure is only semi-automatic because  the HP 8082A  output pulse,  
which is split passively
into 16 pulses, is only capable of driving the 128 channels of one analog box.

\vspace*{1.cm}

\section{The calorimetric ToF card}
\label{tof}
 
\vspace*{1.cm}

The purpose of this board, labelled FEC in
 Fig. \ref{Bloc-diag} (for Front End Card),
is to perform channel by channel: 
\begin{itemize}
  \item
  The on-line timing selection of physics events and the rejection 
  of the proton background originating upstream of the beam interaction point,
  \item
  The shaping and analog summation of the  pulses provided to  the electron 
 ($\sum_{4}$) and background veto ($\sum_{16}$) trigger electronics 
 (Fig.\,\ref{Bloc-diag}).
\end{itemize}
 The inputs to this board are the 16 analog pulses supplied by the
 preamplifiers, via the energy read-out card and the
HERA clock. The outputs provided are five analog sums
($\,4\times\sum_{4}$ and $1\times\sum_{16}$) for the 
electron and total energy triggers and 16 differential
signals (CFD outputs) which are sent to the TDC system.

In the timing branch (Fig. \ref{Bloc-diag}), the pole of the 
preamplifier is removed by
pole-zero compensation (PZC, zero=225~{\rm ns} and pole$\sim$3~{\rm ns}).
In order to prevent CFD oscillations,
 the time constant of the zero (225~{\rm ns}) 
is chosen  slightly larger than the expected maximum value of the preamplifier 
decay time $\tau_{pm}$.

The output signals of the CFD are fed in parallel into the TDC system
(Sect.\ref{tdc})
and the ToF switch. By default the switches are positioned such that 
  shaped pulses of the trigger branch are fed into the AToF (for 
  Anti ToF,i.e. out of time with respect to the beam interaction) branch.
  The switching to a ToF position requires that the leading edge of the 
  CFD output pulse occurs in a time range, called ToF window, covering the 
  energy deposition time of the physics events. The default position AToF 
  is automatically restored 150 ns after the
 AToF$\,\longrightarrow\,$ToF   switching.
  The ToF window is a gate derived for each bunch-crossing from the HERA
  clock.   In order to take into 
account, channel by channel, transit-time  differences of SpaCal pulses and
the HERA clock distribution in the boards, the ToF window position is 
individually tunable by a programmable delay-line
(AD9500 chip). The rear edge of the ToF window is also derived from the HERA
clock and is tunable by a programmable 
delay-line which is, for simplicity, common to the 16 channels of a board. 
This implies that ToF window widths vary from channel to channel, an effect
which is taken into account during time calibration procedures. 
The 8-bit-word programmable delay lines of the ToF windows cover a 
70~{\rm ns} delay range in 270~{\rm ps} steps of the Programmable
Delay Line (PDL).

In the following a more detailed description of the
the constant-fraction discriminator, the trigger shaping, and
the on-line time selection switches is given.

\begin{itemize}

\item The CFD circuit, based on the AD96687 comparator chip, runs in ARC 
(Amplitude Risetime Compensated) mode
with a constant fraction of 20$\%$ and a delay of 4~{\rm ns}. 
No amplitude validation is done in order to
perform the on-line fast switching (Fig.1) within $\sim$~5~{\rm ns}; 
this implies setting the CFD threshold above the noise, as shown below. 
This voltage threshold is tunable by an 8-bit programmable
DAC in the range from -50~{\rm mV} up to +10~{\rm mV} in $\sim 230 \mu$V steps.

\item In the trigger branch, the SpaCal pulse is shaped by a (CR)(RC)$^3$ 
amplifier after PZC (zero=225ns). 
The unipolar shaped pulse has a typical rise time of 20\,ns and returns
to the base line, at a level of 1$\%$, within $\sim$120\,ns.  
The  three integrations of the shaper provide the $\sim$5\,ns delay necessary
for setting the switches (Fig.\,\ref{Bloc-diag}).
\item  The trigger pulse  is fed into a pair of switches, 
one for the AToF line and the other for the ToF one. 
These switches consist of two BSS83 transistors (Philips) chosen for their 
intrinsic delay of $\sim$4\,ns with a small jitter of $\sim$0.5\,ns. 
The two-transistor configuration suppresses fully the spike 
(ref. \cite{Stefan}) induced by the switching.
\end{itemize}

As explained above, during normal data-taking, these switches are activated
(free position)  by the timing  of the energy deposition. 
For calibration purposes, the switch states can be downloaded,
 channel by channel, to permanent AToF, ToF or null position. 
In this last mode,
the channel contribution is removed in both AToF and ToF electronics. 
This configuration, which ensures the possibility to fully test the
trigger electronics, also enables suppression of contributions from noisy 
channels during data-taking.

\begin{figure}[htb]
  \centering
  \begin{minipage}[t]{0.49\textwidth} 
      {\epsfig{file=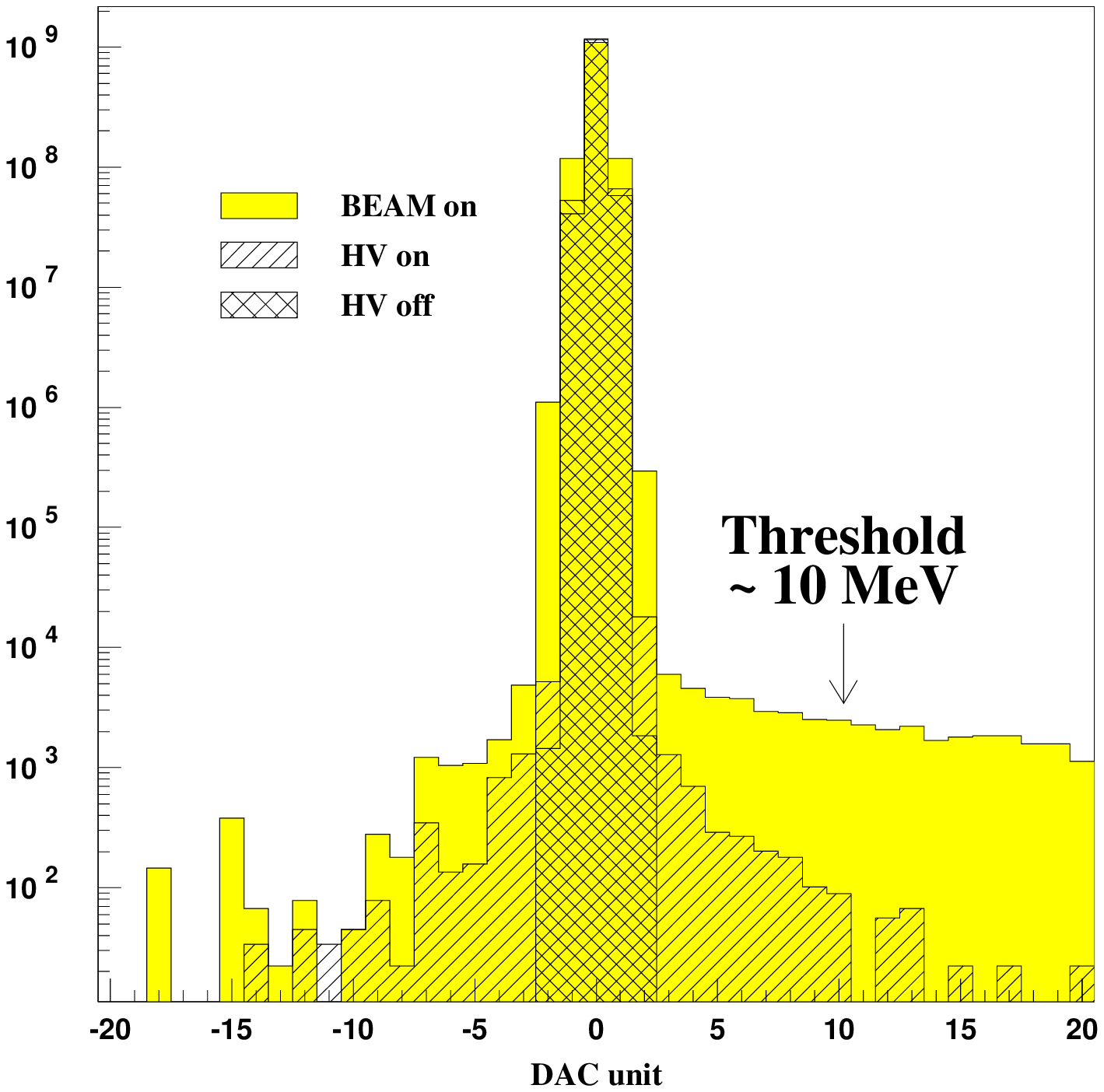,height=7cm,clip=,%     
        bbllx=68pt,bblly=182pt,bburx=496pt,bbury=623pt}}
       \caption{CFD noise of the 128 hadronic channels in DAC units.
        The large tail (BEAM on) is due to small energy deposits by 
        physics events. 
        During data taking, the threshold is set to an energy equivalent 
        value of
        $\sim$~10~{\rm MeV}. The noise RMS of the narrow distribution (HV off)
        is equal to $\sim$\,0.3~{\rm MeV}. The equivalent performance achieved
        with the electromagnetic calorimeter (with PMs of half gain) is 
       $\sim$ 1.5 MeV.}
   \label{cfdfig}                                                        
  \end{minipage}
   \hfill
   \begin{minipage}[t]{0.40\textwidth} 
    \centering
     {\epsfig{file=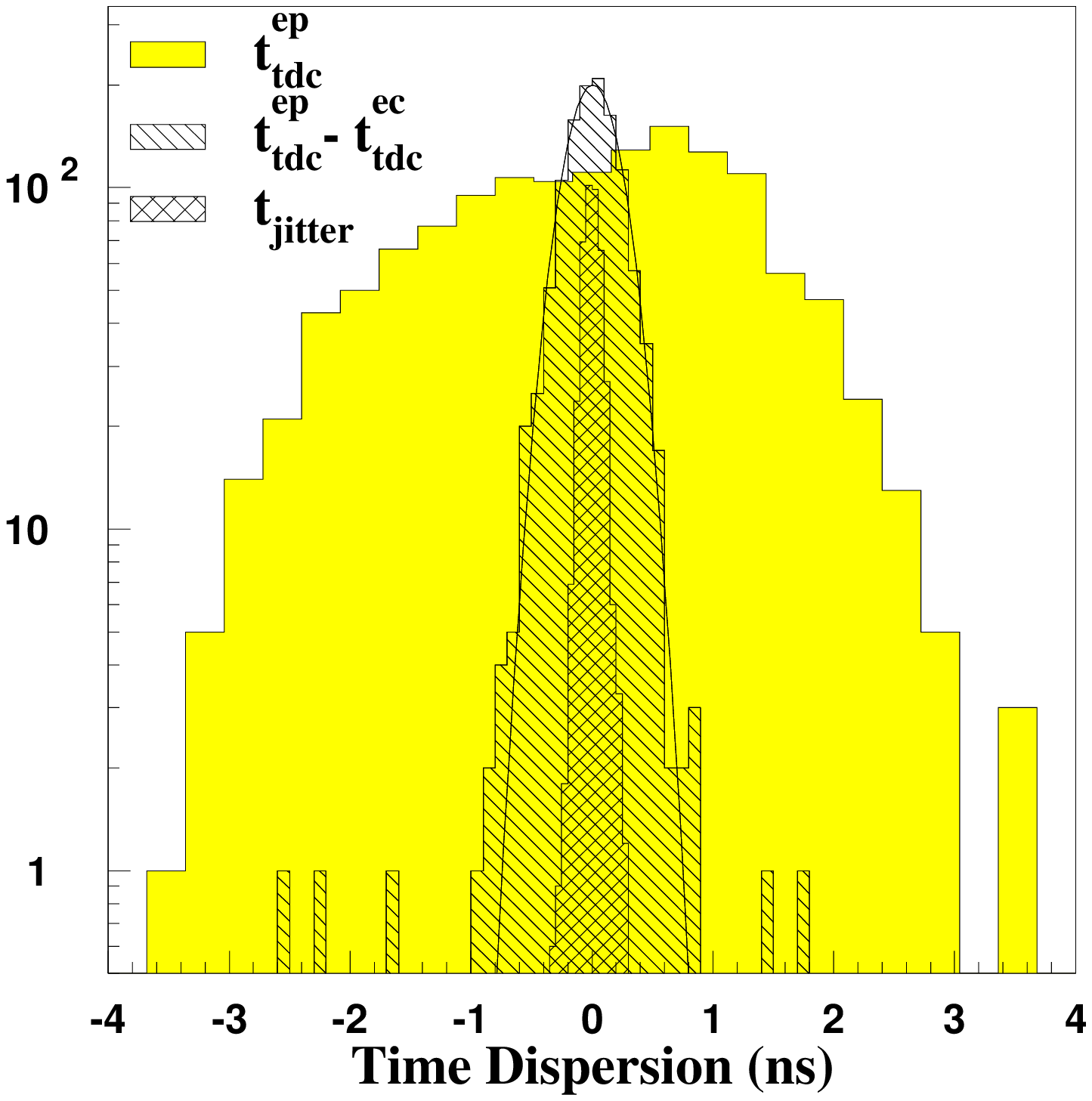,height=7cm,clip=,%  
       bbllx=68pt,bblly=182pt,bburx=496pt,bbury=623pt}}
      \caption{Result of the timing adjustment of the ToF window
         for one analog box (128 channels), using the electronic 
         calibration. Starts and stops are aligned 
         within 270~{\rm ps} of delay step, with a sharpness  
         better than 100~{\rm ps} (narrowest distribution). 
         Refer to text for details.}
  \end{minipage}
  \label{twfig}
\end{figure}

 The noise and timing calibrations of the calorimetric ToF board will now be
discussed.
Fig. (\ref{cfdfig}) shows, as function of the CFD threshold
(in DAC units), the differential output CFD pulse rates measured
by the TDC system (Sect. \ref{tdc}), with particles (beam on), 
without particles (HV on)
and with  photomultiplier high-voltages switched off. Each histogram sums up
the 128 channels of the hadronic calorimeter. The narrower distribution,
which measures the intrinsic electronic noise, is mainly contained 
within one bin and corresponds to a noise 
RMS of $\sim$~67~$\mu$V (0.3~MeV). 
This value, which is approximately 4 times the expected noise contribution 
from the preamplifier, indicates that the transmission of
the high frequency range of  SpaCal pulses is performed 
within a dynamic range of $\sim${15} bits.  A similar value
is obtained for the electromagnetic calorimeter, except for the channels
 of three analog boxes for which the RMS noise is increased up to
200 $\mu$V, the extra noise contribution being induced by
 the low voltage power supplies.
During data taking the CFD thresholds are set at a  value of 
$\sim$~20~{\rm MeV} (10~{\rm MeV} with the hadronic gain factor, see
Fig. \ref{cfdfig}), well above the noise in order to ensure
 the long-term running stability.

The goal of the time calibration is to tune, channel by channel, 
the position of the ToF window.
This is done by setting the switches of each channel
in  turn to the free  position while all other
channels are forced to null position, and then by delaying
for two time settings of the electronic calibration 
pulses, the ToF window in steps of 1 DAC count of 270~{\rm ps},
 equal to the DAC step of the PDL (270 {\rm ps}).
The transition time
of the AToF/ToF switching is recorded using the total energy trigger sums 
(Sect. \ref{trigger}) 
and is related to the corresponding TDC time of the electronic pulses 
(for a more accurate measurement of this
transition time, the calibration system is pulsed five times for each step 
(see \cite{zini})).
From the linear relationship between the TDC time and the PDL unit, 
the setting of the ToF window
is calculated once the energy deposit time t$^{ep}_{tdc}$ corresponding to 
ep collisions has been measured. The start of the ToF window is set 
2~{\rm ns} earlier than t$^{ep}_{tdc}$.

Fig. (\ref{twfig}) shows for the 1192 channels of the 
electromagnetic calorimeter:
 \begin{itemize}
         \item 
         The dispersion of the raw t$^{ep}_{tdc}$ values as measured 
         after one year of data-taking.

         \item 
         The relative time dispersion of t$^{ep}_{tdc}$--t$^{ec}_{tdc}$ 
         between t$^{ep}_{tdc}$ and the
         electronics calibration t$^{ec}_{tdc}$.
         The narrow  RMS width ($\sigma \sim $0.23~{\rm ns}) of the latter
         distribution
         as opposed to the 1.25~{\rm ns} RMS  width of the raw t$^{ep}_{tdc}$
         one indicates that the dispersion of t$^{ep}_{tdc}$
         is mainly generated between the calorimetric ToF board and the TDC 
         system.

        \item 
         The dispersion of the transition time width  from AToF to ToF 
         (and the converse) as measured by the electronic pulses during the 
         ToF window calibration.  This  distribution%
        \footnote{For the display, the bin contents have been downscaled by a 
         factor of .3},
        which is mostly included within one DAC unit (270~{\rm ps}),  
        indicates that the ToF window time jitter ($t_{\rm jitter}$)
        has an RMS value lower than $\sim$0.08~{\rm ns}.
 
 \end{itemize}

 Electronics tests have shown that the timing stability is better than 
   0.08 ns on the few minutes scale. On longer time scales there exist no 
   means to perform this check since long term drifts of the trigger 
   controller and beam interaction time with respect to the HERA clock 
   are larger than 1 ns.

\vspace*{1.cm}
\section{The TDC system }
\label{tdc}

A system of about 1400 TDCs, one per 
photomultiplier, is needed to record the pulse timing in each SpaCal 
channel. This is necessary in order to be able to monitor and adjust 
the timing of the ToF/AToF switching in the trigger, as well as to 
distinguish offline between energy depositions coming from interactions 
rather than from beam-induced background.

Due to the first-level trigger latency, TDC results must be 
pipelined for at least 2.5$\,\mu$s, but such TDCs are 
not available commercially. Therefore, a custom-made system has been
built \cite{eric,nicholls}, based on the commercially available TMC1004
 \cite{TMCpaper1,TMCpaper2}
(Time Memory Cell) developed for use 
on the Superconducting Super Collider. 

Online monitoring was a serious consideration in the design. There
is automatic on-board histogramming of the timing, for different 
types of trigger conditions, in order to allow checking of both the 
calorimeter and the electronics. There are also scalers for measuring 
the CFD output rates for calibration purposes as well as for 
on-line monitoring during data taking.
The TMC chip and the way it is used in 
H1, the architecture and the boards of the TDC system, and the readout
and on-line monitoring are described in the following subsections.
\subsection{The TMC1004 ASIC}

\begin{figure}[htb]
  \begin{center}
{\epsfig{
   file=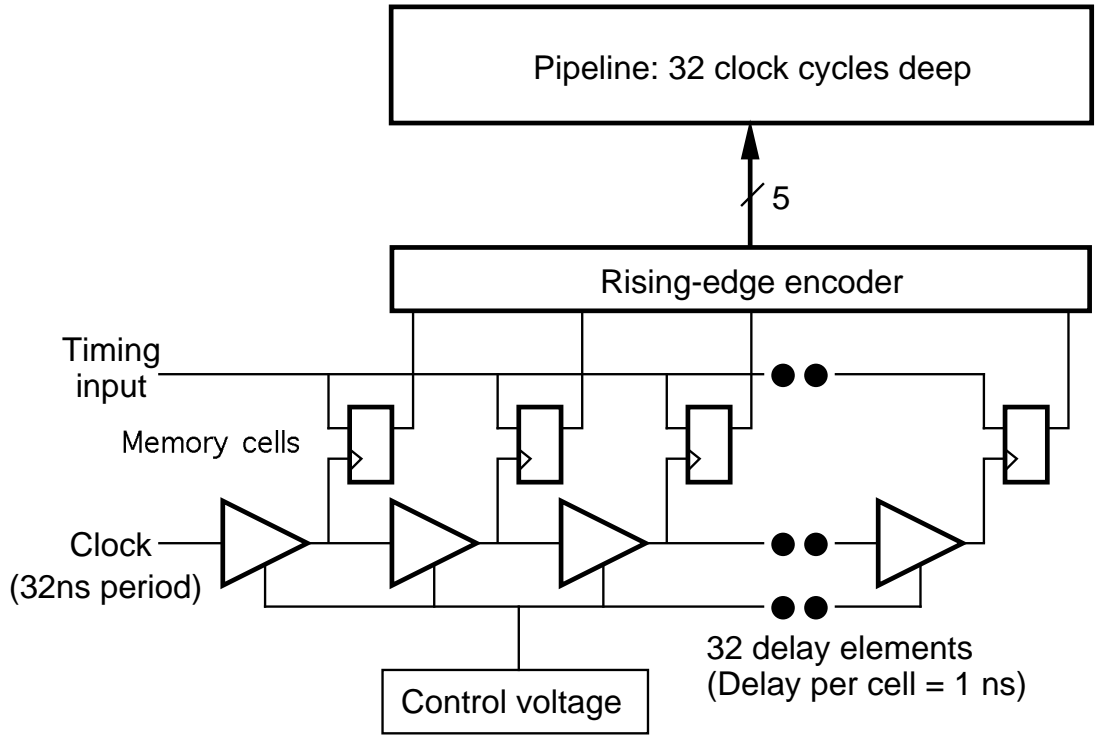,width=0.9\textwidth}}
  \end{center}
  \caption{Conceptual diagram of the TMC1004 chip.}
  \label{TMC}
\end{figure}

A conceptual diagram of the device is shown in Fig.~\ref{TMC}. 
Each ASIC contains four channels, with individual timing inputs and a 
common clock. The digitisation is performed using a chain of 32 delay 
elements controlling the write inputs to a corresponding chain of 
memory cells. The timing input is connected to the data inputs of all 
the memory cells. A clock edge propagates down the chain of delay 
elements, storing the state of the timing inputs in the memory cells 
at intervals given by the delay of the delay elements. The position 
of a zero-to-one transition in the data stored in the memory cells 
indicates the time between the rising edge of the clock and the 
timing input signal. The data are stored in a pipeline, implemented 
as a circular buffer so its depth can be configured to be up to 32 
clock cycles long.

The precision of the TMC1004 alone is specified to be 0.5$\,$ns. The 
bin size, or accuracy of the TDC measurement, can be changed by varying 
the clock frequency, a feature available in the TDC system. The range 
over which the control of the delay is guaranteed is 0.6$\,$ns to 
1.5$\,$ns, corresponding to clock frequencies of 50$\,$MHz and 
21$\,$MHz respectively.

A TMC clock period of 32~{\rm ns}, i.e. covering
only one-third of a bunch crossing, is chosen. This is because the depth of 
the TMC1004 pipelines is only 1$\,\mu$s if data are recorded continuously 
(32$~\times$~32~{\rm ns}), which is not long enough for the H1 trigger 
latency. The solution adopted, in order to avoid the need to extend 
the pipelines with external memories, is to record data from only one
third of a HERA clock period per bunch 
crossing, which extends the length of the 
pipelines to 3$\,\mu$s regardless of the TMC clock period. 
Experience showed that the full base-width of the `interesting' part 
of the bunch crossing could be accommodated in the resulting 
32~ns-wide active window. Thus, the normal situation for running uses
a 31.3~{\rm MHz} clock covering a window of 32~{\rm ns} with a 1.0~{\rm ns}
bin size. However, the TMC clock multiple can be changed from its normal 
value of three to as much as five (52.1~{\rm MHz}, covering 19.2~{\rm ns} 
with a bin size of 0.75~{\rm ns}) or to as little as one (10.4~{\rm MHz}, 
covering the full 96~{\rm ns} bunch crossing with a bin size of 
3.0~{\rm ns}). This trade-off means that precise measurements can be made 
over small windows, and cruder measurements can be made over the 
entire bunch crossing if necessary.

\subsection{TDC system architecture}

The TDC system is contained in two 9U$\,\times\,400\,$mm deep crates with 
standard J1 VMEbus.  The 
VMEbus is used for control and online monitoring, while readout to the 
H1 data acquisition system is via a dedicated bus to the same DSPs as 
the SpaCal energy read-out. 
A custom J2/3 bus using five-row DIN 
connectors allows for rear plug-in of all inputs, and the fanout of 
clock and control signals. The J2/3 bus also provides a separate 
acquisition bus over which data are sent to the readout DSPs.  
The TDC system is integrated into 
the SpaCal electronics as shown in Fig.~\ref{Bloc-diag}.

\begin{figure}[htb]
  \begin{center}
  \mbox{\epsfig{figure=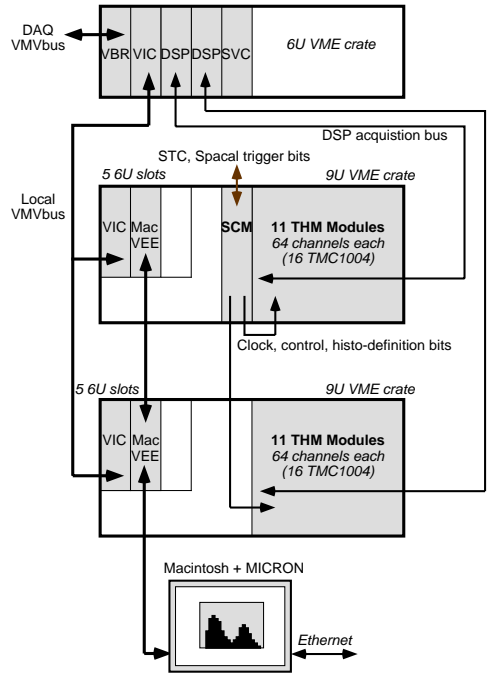,height=9cm}}
  \end{center}
  \caption{Block diagram of the TDC system.}
  \label{TDCsystem}
\end{figure}

The TDC function is performed on 22 TDC and Histogrammer Modules 
(THMs, Fig.7), each with 64 channels and thus 16 TMC1004s. These modules 
receive the output CFD pulses from the front-end electronics 
via 23~{\rm m}-long multi-way 
shielded twisted-pair cables.
The THM includes also a 32-bit scaler per 16 TDC channels for the CFD rate
measurements (Xilinx XC3020). 
Individual channels can be gated on or off, so that these 
scalers can be used either to monitor overall rates in blocks of 16 
channels, or to examine individual channels one-by-one. The latter is 
useful not only in online monitoring, but also for setting the 
threshold of the front-end constant-fraction discriminators to be 
 above the noise. 
 
One System Controller Module (SCM) receives the central HERA 
bunch-crossing clock (10.4~{\rm MHz}) and the Pipeline Enable (PE)
sent by the L1 trigger. It 
multiplies the bunch-crossing clock by a programmable multiple, 
between one and five but normally three, to produce the TMC clock. 
The TMC clock and PE are fanned out to all THMs on 
equal-length cables. The SCM also provides 32-bit scalers for use by 
the monitoring software (e.g. number of bunch crossings, number of 
PEs, etc.). These are implemented in Xilinx XC3042s FPGAs. 

The contents of the TDCs, scalers and histograms are accessible to a 
Macintosh-based C++ monitoring program. The use of a private bus for 
communication with the DSPs allows the Macintosh uninterrupted VME 
access, even during data-taking.

\begin{figure}[h]
  \begin{center}
{\epsfig{file=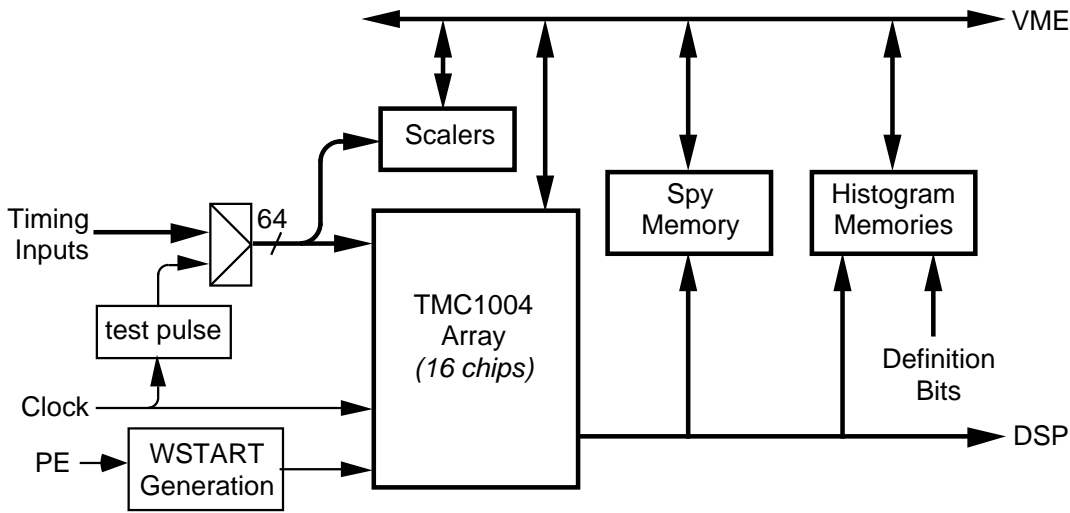,width=0.9\textwidth}}
  \end{center}
  \caption{Block diagram of the THM.}
  \label{THM}
\end{figure}

\subsection{TDC read-out and monitoring}
While the pipelines are enabled, the TMC chips acquire the time of 
arrival of energy with respect to the TMC clock for each channel. 
When a first-level trigger occurs (2.5$\,\mu$s after the bunch crossing 
of interest) the pipelines are frozen, and data from the triggered 
bunch crossing as well as one either side for safety are accessible to 
calorimeter-readout DSPs. Two DSPs situated in a remote 
data-acquisition crate read the data over the local acquisition bus. 
Each DSP reads out one crate of THMs, zero-suppresses the data, and 
places them into a buffer for the event builder.

For online monitoring purposes, automatic histogramming of every 
channel using on-board memories was implemented. This allows the 
monitoring computer to read out filled histograms when it is 
convenient, rather than have to read out many events and build up the 
histograms itself. 

There are actually 16 histograms per channel. Data from a given bunch 
crossing are added to the histogram specified by a set of four 
`definition bits' fanned out from the SCM. The bits are generated from 
the SpaCal trigger bits using a RAM-based lookup table that allows 
specification of arbitrary logical combinations of trigger bits. In 
this way, the significance of the histograms is completely programmable 
and can be changed to allow different investigations to be carried 
out for different types of trigger in the SpaCal.

There is one histogrammer unit per 16 channels. Although only one of 
these channels can be selected for histogramming at a time, all 16 can 
be stepped through in sequence automatically. The histograms are stored 
in RAM, requiring 1024 32-bit words per channel. This gives a total of 
5.5~Mbytes of histogram RAM in the TDC system. To histogram the data 
from a single bunch crossing takes eight bunch crossings, during which 
the histogrammer is insensitive. With zero-suppression this is 
acceptable, given the low occupancy of each channel. 

These histograms, along with the scalers on the SCM and THMs, are 
accessible via VMEbus to the monitoring software for online checks 
and expert diagnoses. 

\newpage
   
\section{Electron and Total Energy Triggers }
\label{trigger}

  \subsection{Inclusive electron trigger}
  \label{IET}
 
     Since the current jet of a low Bjorken $x$ deep-inelastic event is boosted
in the backward direction, a total energy sum cannot be used at the first level
trigger to distinguish deep inelastic scattering events from photoproduction 
events  ({\sl Q$^{2}$} $\ll$ 1~{\rm GeV$^{2}$}).
We therefore apply a simple inclusive electron trigger condition that there 
is at least \textit{one} trigger tower (4$\times$4 SpaCal towers) energy 
greater than a threshold. 
More precisely, this inclusive electron trigger (IET) is
 designed to compare the deposited energy
    inside a group of 4$\times$4 SpaCal towers with each of 3 thresholds.
    A global `OR' of the digital outputs for each threshold is performed, 
    the result of 
    the three `OR's being sent to the H1 first-level central trigger logic. 
        The trigger tower size is chosen 
    such as to ensure the transverse containment of the electromagnetic shower
    at any scattering angle $(155^0<\Theta<177^0)$.

\begin{figure}[htbp]
  \begin{center}                                                         
  {\epsfig{file=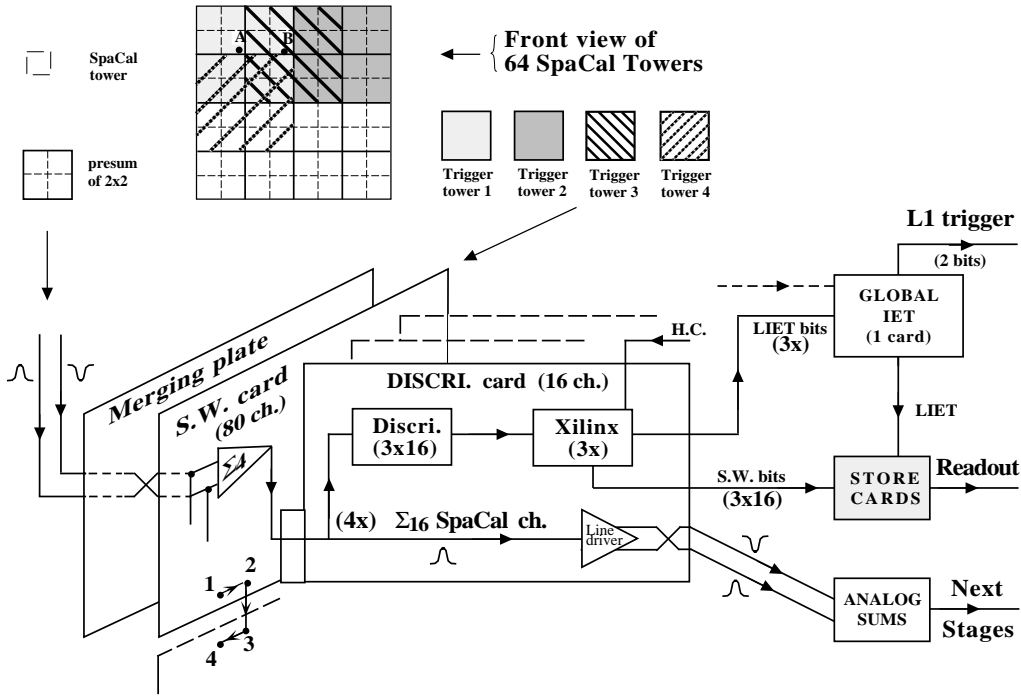,width=0.9\textwidth}}
  \end{center}
  \caption{\small{ General layout of the IET}} 
  \label{iet}
 \end{figure}

    The sliding trigger towers are formed  as shown at the top of 
    Fig. \ref{iet}. From hardwired presums
    $\Sigma_{4}$ of the ToF-timed signals from 
    2$\times$2 SpaCal towers (done in
    the calorimetric ToF card described in Sect. \ref{tof})
    the content of all $\Sigma_{4}$ presums (i.e. the overlaying sums of 16
    SpaCal towers) is 
    calculated (Fig. \ref{iet}). 
    In this way, even when the impact point is at the border of 
    two presums (point B in Fig. \ref{iet}), the full deposited energy is 
    recovered in 
    the trigger tower 3 (Fig. \ref{iet}). 
    As shown in this figure, the sliding summation 
    is performed in both directions, x and y.

      A schematic view of the electronics is shown in Fig. \ref{iet}.
    The pulses (fixed presums) supplied by the calorimetric card 
    are sent to the trigger electronics on  shielded multi twisted-pair cables.
   The merging plate allows each presum to be individually mapped to the
   correct destination in order to do the sliding-sum calculation.
    As sketched in Fig. \ref{iet}, a basic trigger crate consists of a Sliding 
    Window (SW) card (80~channels) acting as a backplane, and Discriminator 
    Cards (5$\times$)
    processing 16 channels each. The full trigger is contained in 5 such 
    (3U $\times$~350~{\rm mm} deep) crates.
    The main features of these two cards are:

\begin{itemize}
\item[$\bullet$]
      {\it Sliding Window Card}: this is an 8-layer printed-circuit 
      board,  two layers
      used as voltage supply planes $(\pm6~\mathrm{V})$ and the others for 
      the distribution 
      of the analog pulses.
      According to the sliding window scheme (Fig.~\ref{iet}), a differential
      presum pulse must be sent to 4 line-receivers and each line-receiver 
      processes
      4 presum pulses (acting then as summing amplifier). 
      The receiver/summing amplifier (OPA620 from Burr-Brown) is the
      only active component of this card. Receiver outputs corresponding to 
      trigger towers
      are gathered per group of 16 and sent to the discriminator card via a 
      two-row DIN connector. The latter enables the fast signals to be fed 
      with a minimal crosstalk of about 0.5 \%.
      
\item[$\bullet$]             
       {\it Discriminator Card}: The functions of this board are:
\begin{enumerate}
      \item      
           To compare (AD790JR discriminators) the trigger tower pulse height 
           with three thresholds.
      \item      
           To supply, per threshold, a local `OR' 
           (named LIET bit for Local IET) of the 16 discriminated pulses.
           These are latched and synchronized with the HERA clock (HC) 
           inside a Xilinx (XC 3020) which performs this `OR'. The individual 
           bits (SW bits), corresponding to detailed trigger information 
           used for debugging or analysis, are read out via `Store Cards'
           (\cite{Orsay}).
       \item     
           To pick-up the analog information of the 4 adjacent trigger 
           towers and to drive them differentially to the summing electronics, 
           as described below.
                 
\end{enumerate}       
\end{itemize}
            As shown in Fig. \ref{iet}, LIET bits are sent to a single card 
            (named `global IET'). This board 
            performs the `OR' of the $3(thresholds)\times$25 LIET bits  and 
            supplies the encoded result to the central L1 trigger.
              
         The main performance features  of this trigger are as follows:       
         the RMS output noise of a receiver/summing amplifier is equal 
         typically to 
         $\sigma=0.3$~{\rm mV} which is equivalent to $\sim$\,8~{\rm MeV}. The
         cross-talk signal, generated inside the SW card, between trigger
         tower pulses, is of the order of ${\sim\,10^{-3}}$,
         which is negligible. The amplitude dispersion $\sigma$ at the output
         of the SW cards has been measured to be $\sim\,1.1\%$ 
         (\cite{Stefan}); 
         no significant spatial pattern in the response of these cards was 
         found, especially when the presums are distributed to two SW cards 
         as indicated in Fig. \ref{iet}.

  \subsection{Total energy trigger}
  \label{totenergy} 
\begin{figure}[htbp]
 \begin{center}                                                         
{\epsfig{file=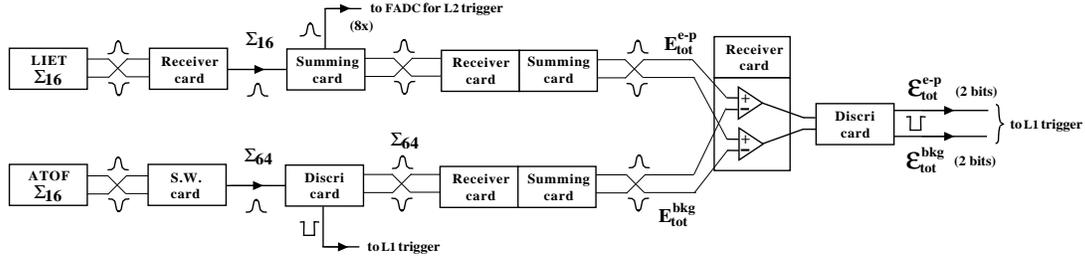,width=0.9\textwidth}}
 \end{center}
 \caption{\small{ General layout of total energy trigger}} 
 \label{etot}
 \end{figure}

  In this section we describe the electronics (Fig. \ref{etot}) used to 
  generate the trigger elements $E_{tot}$ of the total energy for both 
  in time (ToF or ep) and out of time (AToF or background) events. 
  They are based on the IET electronics and on two new boards, 
  the Receiver card (80~channels)
  and the Summing card (16~channels) which can be arranged in a crate 
  in a similar way as 
  the SW and discriminator cards respectively. The receiver card is a 
  simplified version of the SW card, the OPA620 being mounted 
  in the usual line 
  receiver configuration.
  The Summing card performs the analog sums  $\sum_{16}$ or 
  $2\times\sum_8$ or $4\times\sum_4 $ input channels; the 
  option of the sums is set inside the card by jumpers. 
  This allows great flexibility for doing sums according to a given mapping.
          
          The 77 input channels of the summing electronics are processed
    differently 
   for ep and background events, as can be seen in Fig. \ref{etot}:
\begin{itemize}
\item   
   For ep events, the $\sum_{16}$ SpaCal channels
    with ToF signals supplied by the LIET cards are
    summed into 8 `big sums'  which correspond to large regions of 
    the calorimeter
    (4 quarters of an inner square of 16 $\times$ 16 SpaCal
    towers and 4 quarters of the remaining outer shell). 
    The 8 `big sums' are fed into FADCs for the L2 
    trigger and, in parallel, to a final sum operation which provides the 
    total energy E$_{tot}^{ep}$.
                        
\item     
    For events outside of the ToF window (AToF events), 
    the $\sum_{16}$ SpaCal channels with AToF signals 
    supplied by the calorimetric ToF 
    cards are processed by an electron trigger crate (Fig.~\ref{etot}),  
    using only
    one threshold for the energy comparison.
    The `OR' of the 5 local AToF bits is sent to the 
    L1 trigger. The purpose of the sliding sum board for AToF energies
    is to recover, at the output of the discriminator cards,
    the complete analog information for the total energy E$_{tot}^{bkg}$
    summation. 
    The latter is done in the same way as for the ep events (Fig.~\ref{etot}).

\end{itemize}

    The right part of Fig.~\ref{etot} shows that  the quantities 
     E$_{tot}^{ep}$--E$_{tot}^{bkg}$ and E$_{tot}^{bkg}$--E$_{tot}^{ep}$
     are discriminated in place 
     of the raw total energies E$_{tot}$. This energy comparison is made to 
     prevent  spurious triggers which are attributable to the following :
\begin{enumerate}          
\item          
    Inside the calorimetric ToF board, the slewing of the CFD outputs 
    for small energy deposits  
    ($\sim$   CFD threshold value) can direct a non-negligible fraction of the 
    total energy of the event towards the trigger branch with 
    inappropriate timing
    (AToF instead of ToF or vice-versa).
\item  
    For an ep event, all energy deposits below the CFD threshold are not
    timed; they are sent
    by default to the AToF branch and can lead to a trigger veto.
\item
    High frequency cross-talk inside the energy read-out and the ToF cards,
    summed
    over  the 16 neighbouring channels.
\end{enumerate}              
    The first two effects dominate at low energy and the last one at high 
    energy (4\% of AToF to ToF
    above 25~{\rm GeV} and 2\% ToF to AToF above 25~{\rm GeV}).    

    As can been seen in Fig.~\ref{etot}, the subtraction is performed by 
    feeding the same polarity of the two differential pulses into a line 
    receiver amplifier of a Receiver card. The result of this analog 
    subtraction is compared
    with two thresholds of a discriminator card.  

    Although H1 has not yet run with the SpaCal energy thresholds set to 
    values below 
    600~{\rm  MeV}, the electronic noise alone allows 100~{\rm MeV} 
    thresholds for the IET,
    well below the 300~{\rm MeV} level for a minimum ionising particle 
    ({\it mip}). 
 
\vspace*{1.cm}

\section{Performance at HERA }
\label{perfo}

\begin{figure}[htb]   
 \begin{minipage}[t]{0.45\textwidth} 
 \centering
{\epsfig{file=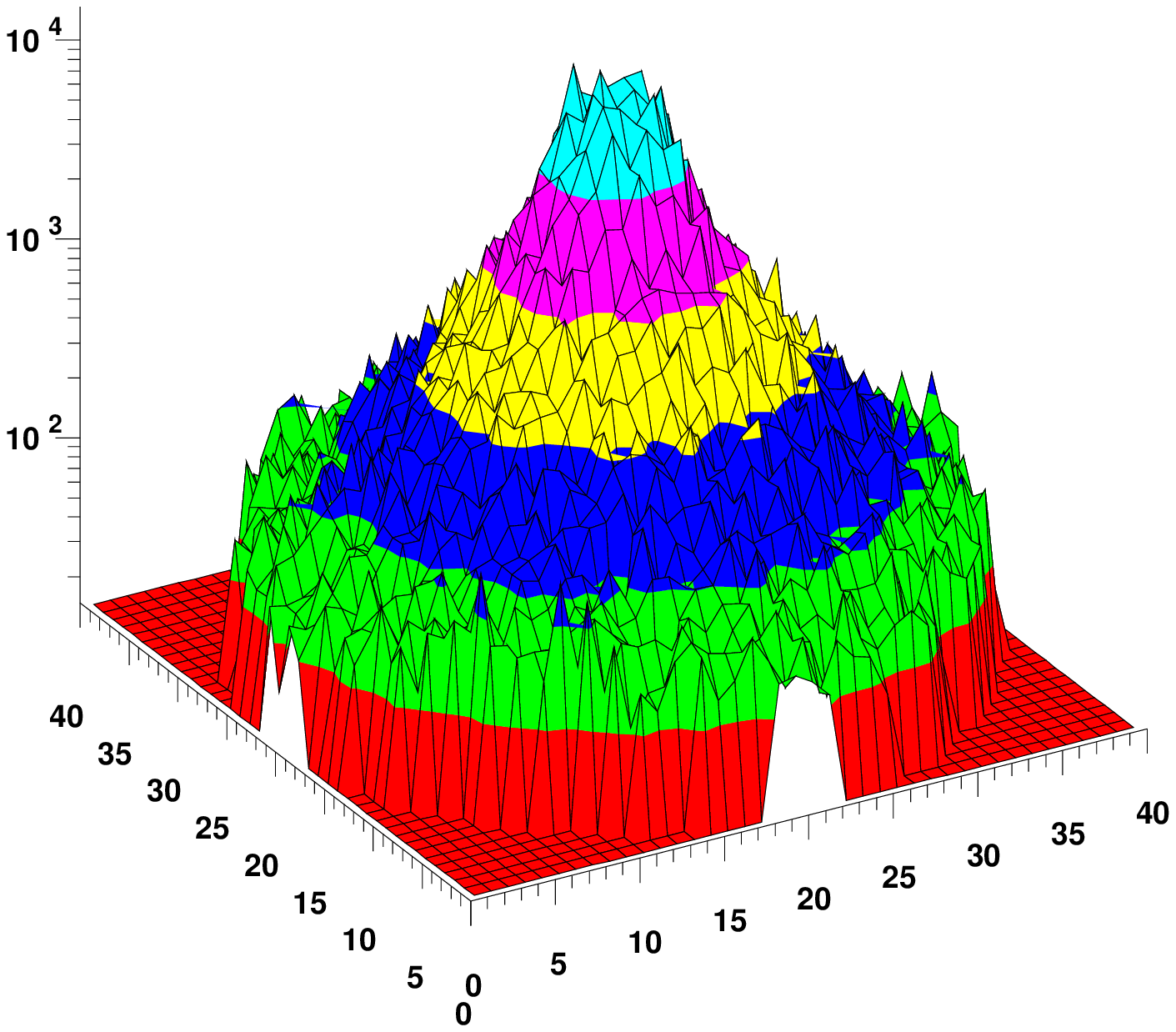,height=6.0cm%
 ,bbllx=78pt%
  ,bblly=183pt%
   ,bburx=494pt%
     ,bbury=574pt%
 ,clip=}}
\begin{rotate}{-38}
\put(-160,-80){Cell number}
\end{rotate}
\begin{rotate}{14}
\put(-80,20){Cell number}
\end{rotate}
\put(-205,150){Hz}
  \caption{2-dimensional distribution of the counting rates (Hz) of
    the 1192 cells of the SpaCal EM calorimeter, measured on-line by
    the TDC system.}
  \label{rates}
\end{minipage}
 \hfill
\begin{minipage}[t]{0.40\textwidth} 
 \centering
{\epsfig{file=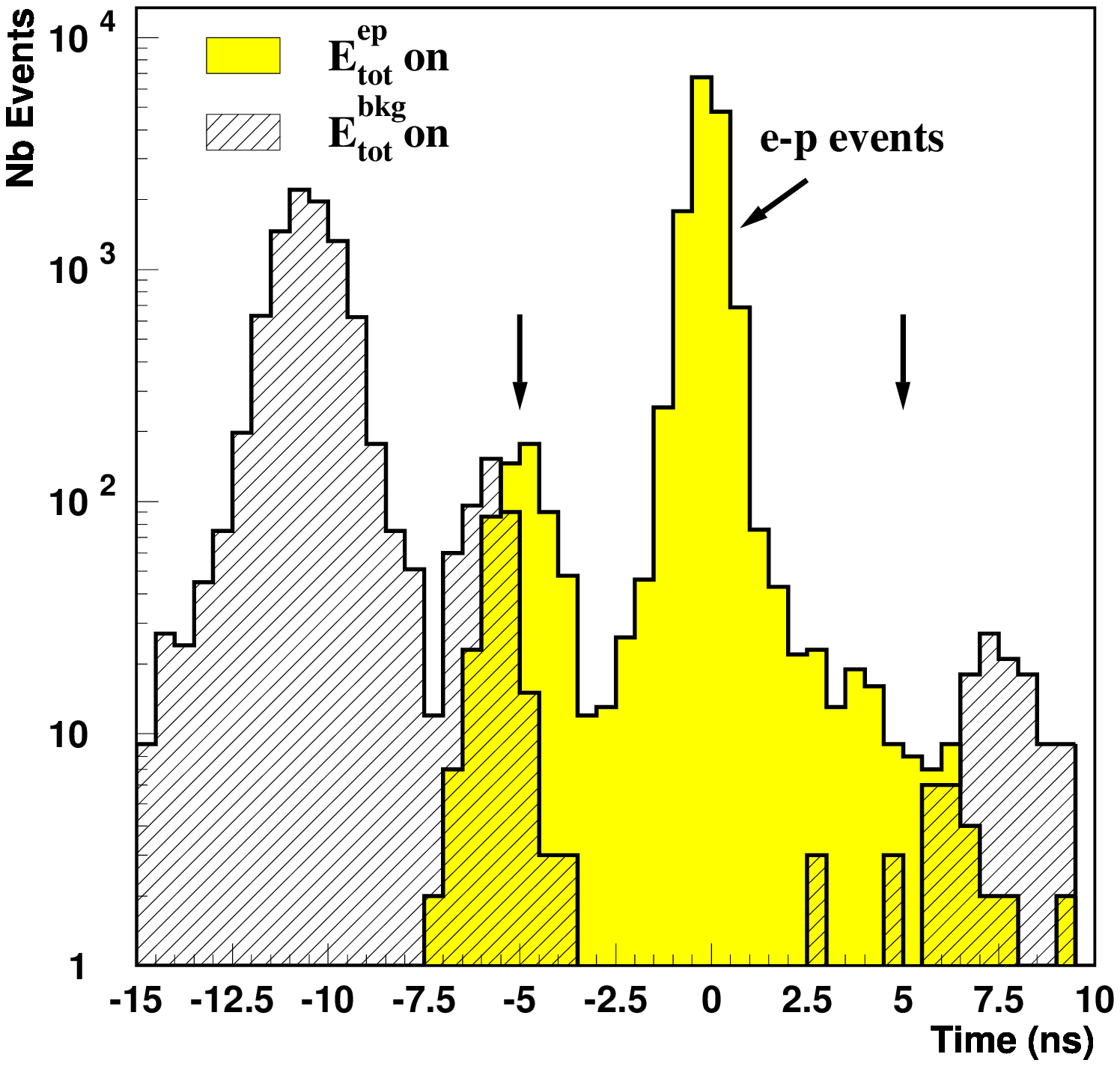,height=6.0cm%
 ,bbllx=26pt%
  ,bblly=166pt%
   ,bburx=525pt%
     ,bbury=648pt%
 ,clip=}}
  \caption{Corrected time distribution (see text) of the energy
    deposit in the SpaCal EM.
  The $p$-background events  seen here are introduced
  in the H1 first level trigger prescaled by a factor $\ge$ 1000 to monitor the
  performance of the various non-SpaCal triggers. 
  The time resolution achieved is $\sigma\,\sim\,$0.35~{\rm ns}.}
  \label{fm2}
\end{minipage}
\end{figure}

During normal running at HERA, the rate scalers (Sect. 6)
are used in a mode
where they cycle through all the SpaCal channels. 
Fig.~\ref{rates} shows such a rate distribution;
 X and Y are the SpaCal tower numbers. This on-line histogramming, as well as 
the timing plots display of individual channels (not shown),
have proven to be a valuable diagnostic 
of both the correct behaviour of the SpaCal and of the backgrounds 
due to the HERA beams. 
The rate distribution (Fig.~\ref{rates}) is peaked around the beam pipe 
at a value of about 15\,kHz. This is  due to the 
proton--gas collisions, physics events accounting for only a small fraction 
($\sim$\,10$^{-3}$) of this distribution. 

Studies of the performance of the SpaCal triggers can be
achieved by recording events without the  on-line rejection
maintained by higher-level triggers  during normal H1 data-taking.
Fig.~\ref{fm2} shows a time distribution 
of events recorded during such a  special run. 
The earlier peak is due to background energy deposits, while  the 
second  one corresponds to  ep physics events, the small one in between 
coming from 
background due to the proton beam structure (proton satellite bunch).

In Fig.~\ref{fm2}, the time of the event has been calculated off-line
in the following steps:
\begin{enumerate}
     \item
      For the i-th cell, the raw time is corrected (t$^{corr}_{i}$) from
      the offset  t$^{ep}_{tdc}$ (sect.\ref{tof}) and the slew effect of
      the CFD,
     \item
      The event time is defined as the mean weighted time:
\begin{equation}
 t = \frac{\Sigma_{i} E_{i} t_i^{corr}}{\Sigma_{i} E_{i}}
\end{equation}
 where $E_i$ is the energy deposited in the $i$-th cell. This linear weighting 
has the advantage of lowering
the contribution of energy deposits which are approximately equal
to the CFD threshold ($\sim$\,40~MeV); for these low energies, 
the time is measured with a jitter of a few ns due to the noise
 fluctuation at the input
of the CFD
      \item
      The ep distribution mean value is taken as the time origin. 
Due to the dispersion of the offsets  $t^{ep}_{tdc}$,
  the effective histogramming time window is reduced from 32~{\rm ns}
 to  $\sim\,$25~{\rm ns}. This window is still wide enough to contain
 both ToF events and proton beam background.
\end{enumerate}

The two peaks of Fig.~\ref{fm2} are separated by $\sim$ 10~{\rm ns},
as expected. The RMS width of the physics events peak 
is a direct measurement of the time resolution achieved by the 
SpaCal calorimeter and the electronics because the arrival-time dispersion
of physics events in the calorimeter is negligible.
A gaussian fit to data from events triggered by the total energy trigger
$\mathcal{E}$$^{ep}_{tot}$ (see Fig.10) with an energy threshold of
$\sim600$~{\rm MeV} yields a time resolution value  
$\sigma \simeq 0.35$~{\rm ns}, dominated by TDC resolution.
 The width of the background peak
is broader ($\sigma \simeq 0.7$~{\rm ns} after subtraction of the 
0.35~{\rm ns} resolution) and is compatible with 
the proton bunch length measured, for instance,  by the H1
 central tracker detector ($\sigma \simeq 0.6$~{\rm ns}).
When including events triggered by lower energies, the resolution
degrades from 0.35~{\rm ns} at 600~{\rm MeV} to 0.8~{\rm ns}
 for 80~{\rm MeV} (30~{\rm MeV}) in 
the electromagnetic (hadronic) section. 

As can be seen in Fig. 12, besides the two main peaks discussed
above, there is a continuous time distribution 
of energy deposit. These events are related
 to background collisions
of a proton satellite bunch with the residual gas. This satellite 
bunch is delayed with respect to the main proton bunch and varies in
shape and intensity for different fills of the HERA machine.

These satellite background events can be used to study in-situ the transition
 AToF$\,\leftrightarrow\,$ToF performed by the calorimetric 
ToF board. The shaded histogram includes events triggered by the 
total energy trigger $\mathcal{E}$$^{ep}_{tot}$  
with an energy threshold of $\sim600$~{\rm MeV}, 
while the hatched histogram corresponds to events which would have been 
rejected by the veto $\mathcal{E}$$^{bkg}_{tot}$ at the same threshold
value
(see Fig.10); due to the energy subtraction performed at the
last electronics stage,
no events are found simultaneously with the $\mathcal{E}$$^{ep}_{tot}$ and
 $\mathcal{E}$$^{bkg}_{tot}$ trigger bits ON.

The spread of the events selected as ``ep'' events corresponds to
 the average ToF window  width ($\sim\,$10~{\rm ns}) indicated by the 
 vertical arrows.
  The transition AToF$\,\leftrightarrow\,$ToF occurs within
 $\sim\,$0.5~{\rm ns} RMS. The width of this transition region is 
 compatible with the 
 ToF window calibration error, dominated by the TDC resolution (and not
 by the slew time because these events are from high energy background).
 At energies around 1 GeV the transition would be
 broader due to slew effects.

 The two overlap events located at 3~{\rm ns} and 4.5~{\rm ns} (Fig. 12)   
are easily identified from the  off-line TDC information, as 
pile-up events of two consecutive bunches and are thus 
rejected.
 
  As a conclusion about this on-line time-of-flight selection,
it can be remarked that: 
\begin{itemize}
  \item
            Events from the main proton peak are entirely rejected 
            (about $10^8$
            events). The remaining background due to the satellite bunches
            is of the order of $10^{-5}$ of the main proton background. 
            It can be
            reduced to the level of $10^{-6}$ by an off-line timing
            analysis. A further factor 10 can be gained by using the 
            hadronic section of the SpaCal \cite {zini}.

  \item
            For the 600 MeV cut on the total energy  there is no loss of
            physics events. Below this energy threshold, the importance 
            of event topologies requires a special treatment, as described
            in \cite {zini}.
 
\end{itemize}

\begin{figure}[htb]
  \begin{center}
{\epsfig{file=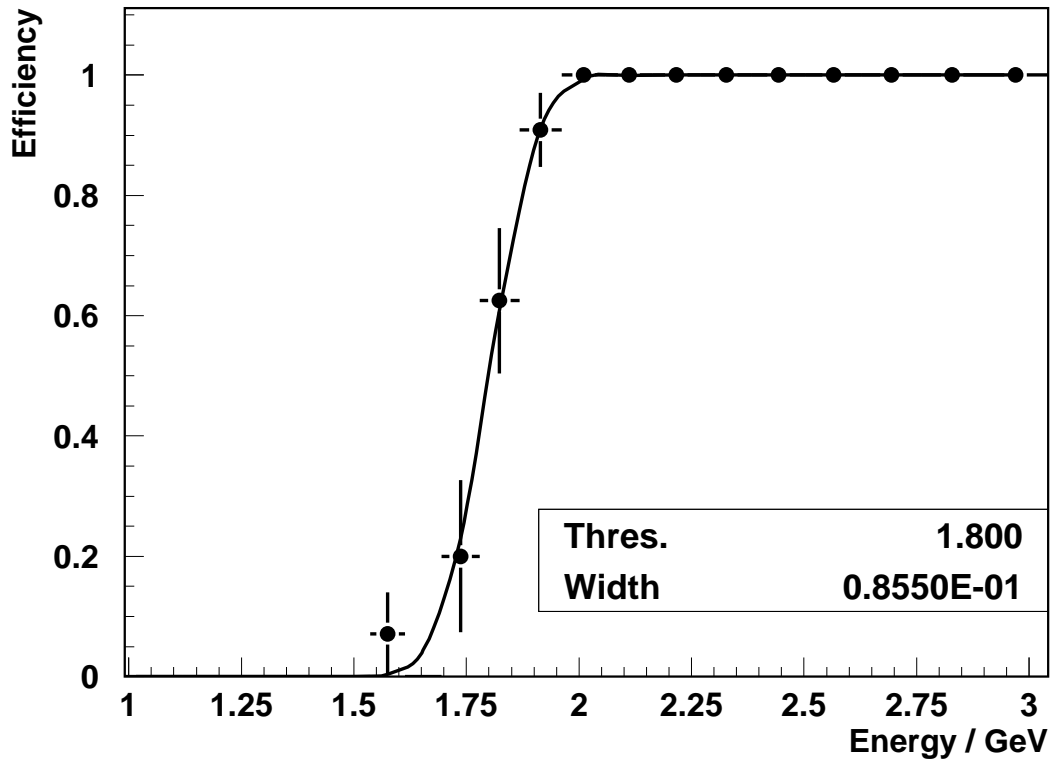,width=0.9\textwidth}}
  \end{center}
  \caption{Combined efficiency of the time-of-flight and electron
    trigger (IET) as a function of the deposited energy inside one
    sliding sum of the IET trigger.}
  \label{ieteffi}
\end{figure}

Fig. 13 shows the combined efficiency of the time-of-flight selection
 and the electron trigger IET, for one sliding sum. This curve is obtained
from the H1 data according to the following procedure: 
\begin{enumerate}
      \item 
       Events triggered without a SpaCal energy requirement and with a
       reconstructed time in the interval of $\pm\,2$~{\rm ns} 
      (Fig.~\ref{fm2}) are selected, 
      \item  
       The deposited energy inside the trigger tower (Sect. \ref{IET})
       is calculated from the energy deposited in the SpaCal towers
       belonging to the trigger tower, 
      \item  
       Early or late pile-up events tagged by the $\pm$~1 bunch-crossing
       TDC information are rejected.
\end{enumerate}
During normal data taking,
 threshold values of the IET are  set approximately to
0.5~{\rm GeV}, 1.8~{\rm GeV} and 5~{\rm GeV} respectively. 
Fig.~\ref{ieteffi} shows that an efficiency of
100$\%$ is currently achieved above 2~{\rm GeV} by the second threshold, 
used mainly to trigger deep inelastic scattering events; a similar 
performance has been shown (ref. \cite{como}) for the lowest threshold
where an efficiency of 100 \% is reached at 600 MeV. The width of the
threshold (Fig.~\ref{ieteffi}) is found equal to $\sigma\sim\,$85~{\rm MeV}. 
This value is $\sim\,$10 times greater than is expected from the total
noise ($\sim\,$8~{\rm MeV}) of the analog sum (Sect.\ref{totenergy}). 
This discrepancy is
due to the variation, from channel to channel, of the charge  preamplification
gain  discussed in  Sect. \ref{bases}. According to the shaping performed in
the next stages, this gain variation is more or less washed out;
the longer shaping time of the readout branch implies that the shaped
pulse amplitude is less sensitive to this gain variation than the one in the 
trigger branch. As the PM high voltages are tuned in order to balance the
response of the channels in the readout branch, the
reconstructed width of the trigger threshold is larger than what could be 
expected only from noise considerations. This gain effect leads to a
threshold width proportional to the energy threshold, but is not a
limitation for triggering with a high efficiency over the full energy range.

\section{Summary}

 The electronics associated with the H1 SpaCal
 lead/scintillating-fibre calorimeters, in operation at HERA since
 1995 has been described. The full electronics was specially 
 developed to have the following
 functionality:
 \begin{itemize}
 \item Wideband ($f \le\,$200~{\rm MHz}) low noise ($\le1$~{\rm MeV})
   preamplification which uses the photomultiplier capacitance for
   charge integration. This preamplifier is, in principle, very
   general and can be applied to any low-capacitance detector.
\item For the readout branch, energy and time information for each
  cell is provided. In particular, the time information has
  proved to be very powerful for the off-line rejection of residual
  background or pile-up events. A time resolution of 0.8~{\rm ns} is obtained
  for SpaCal total energies 
  above 80~{\rm MeV} (30~{\rm MeV}) for the electromagnetic (hadronic) wheel.
\item For the trigger branch, the calorimeter time-of-flight is able
  to reject proton background events with a time resolution of
  0.5~{\rm ns} RMS.
  The low-noise electron trigger, based on analog sliding sums, running at
  the HERA frequency (10~{\rm MHz}) enables the use of 100\% efficient 
  triggers with a cluster  energy threshold as low as 0.5~{\rm GeV}.
\item The low noise performance permits analog summation of
  the 1192 channels of the electromagnetic calorimeter. A minimum-bias
  total-energy trigger with a threshold of $\sim\,$0.6~{\rm GeV} is
  currently used, yielding an out-of-time background rejection
  greater than 10$^6$ with no measurable overefficiency.
  This total-energy sum, together with the
  programmable  switches of the calorimetric ToF board, are very
  powerful tools for the complete tuning and testing, channel by
  channel, of the electron and calorimetric triggers.

 \end{itemize}

 \section{ Acknowledgements}
 The members of the H1 SpaCal group extend their warm thanks to the H1 
 collaboration for
 its stimulating motivation and constant support, and acknowledge the
 outstanding commitment of the many engineers and technicians from the
 different institutes. The authors thank C. de La Taille for his
 invaluable help in the design of the preamplifier. The group also 
 wishes to thank the
 DESY directorate for the hospitality extended to the non-DESY members.


\newpage

\begin{thebibliography}{99}


\bibitem{upgr93} H1 Collaboration, Technical Proposal to Upgrade the Backward 
 Scattering Region of the H1 Detector, DESY PRC 93/02.
\bibitem{spac96} T.~Nicholls \textit{et al.}, H1 SpaCal group, Nucl. Instr. 
 and Meth. {\bf A374} (1996) 149.
\bibitem{spac97} R.D.~Appuhn \textit{et al.}, H1 SpaCal group, Nucl. Inst. 
 and Meth.{\bf A386} (1997) 397.
\bibitem{hadr96} R.D.~Appuhn \textit{et al.}, H1 SpaCal group, DESY 96-013.
\bibitem{calor97} E.~Tzamariudaki, VII Int. Conf. on Calorimetry in HEP, 
 Tucson, AZ, USA (1997).
\bibitem{bemc96} J.~B\'an \textit{et al.}, Nucl. Instr. and Meth. {\bf A372} 
 (1996) 399.
\bibitem{gore96} I.~Gorelov,  VI Int. Conf. on Calorimetry in HEP, Frascati, 
 Rome, Italy (1996),
Frascati Physics series vol.VI, pp 225-236.
\bibitem{dago94} S.~Dagoret \textit{et al.}, Nucl. Instr. and Meth. 
 {\bf A346} (1994) 137.
\bibitem{phot94} J.~Janoth  \textit{et al.}, Nucl. Instr. and Meth. 
 {\bf A350} (1994) 221.
\bibitem{spacpm} R.D.~Appuhn \textit{et al.}, H1 SpaCal group, Nucl.Inst. 
 and Meth. {\bf A404}
    (1998) 265.
\bibitem{rad} V.~Radeka, IEEE Trans. Nucl. Sci. {\bf NS 21} (1974) 51.
\bibitem{taille}See e.g. C.~de La Taille, Th\`ese de Doctorat (in French), 
 Ecole Polytechnique, December 1989.
\bibitem{secre} SECRE Composants, Paris, France.
\bibitem{LAL_seq} R.~Bernier \textit{et al.}, H1 internal note H1-07/92-237.
\bibitem{Lar}  B.~Andrieu \textit {et al.}, H1 Calorimeter group, Nucl. Instr.
 and Meth. {\bf A336} (1993) 460.
\bibitem{H1nim}
I.~Abt \textit {et al.}, H1 Collaboration, DESY report H1-96-01 , 
DESY, Hamburg (1996) and Nucl. Instr. and Meth. {\bf A386} (1997) 310
and ibid. {\bf A386} (1997) 348.
\bibitem{LAL_cal} R.~Bernier \textit {et al.}, H1 internal note H1-04/92-219.
\bibitem{Stefan} S.~Spielmann, Th\`ese de Doctorat (in French), Ecole
  Polytechnique, July 1996. 
\bibitem{zini} P.~Zini, Th\`ese de Doctorat (in French), Universit\'e de 
 Paris VI, July 1998.
\bibitem{eric} E.~Eisenhandler \textit{et al.}, IEEE Trans. Nucl. Sci. 
 {\bf 42} (1995) 688.
\bibitem{nicholls} T.~Nicholls, Ph.D. Thesis, The University of Birmingham,
 April 1997.
\bibitem{TMCpaper1} Y.~Arai \textit{et al.}, IEEE Jour. of Solid State
 Circuits {\bf 27} (1992) 359.
\bibitem{TMCpaper2} Y.~Arai \textit{et al.}, IEEE Trans. Nucl. Sci.
 {\bf 39} (1992) 784.
\bibitem{Orsay} C.~Beigbeider \textit{et al.}, H1 internal notes H1-10/92-242
  and H1-02/93-269.
\bibitem{como} F.~Moreau,  V Int. Conf. on Advanced Technology and
  Particle Physics, Como (Italy) 1996, Nucl. Phys. B (Proc. Suppl.), 132.
  {\bf 61B} (1998).
\end{thebibliography}
\end{document}